\documentclass{article}

\usepackage{arxiv}

\usepackage[utf8]{inputenc}
\usepackage[T1]{fontenc}
\usepackage{hyperref}
\usepackage{url}
\usepackage{booktabs}
\usepackage{amsfonts}
\usepackage{amsmath,amssymb}
\usepackage{nicefrac}
\usepackage{microtype}
\usepackage{cleveref}
\usepackage{graphicx}
\usepackage{natbib}
\usepackage{doi}
\usepackage{mathrsfs}
\usepackage{makecell}
\usepackage{xcolor}
\usepackage{tabularx}
\usepackage{multirow}
\usepackage{tikz}
\usepackage{pgfplots}
\usepackage{caption}
\usepackage{subcaption}
\usepackage{stfloats}
\pgfplotsset{compat=1.18}

\DeclareUnicodeCharacter{2248}{\ensuremath{\approx}}

\title{Bridging Through Absence: How Comeback Researchers Bridge Knowledge Gaps Through Structural Re-emergence}

\author{
  Somyajit Chakraborty\\
  University College Cork, Cork, Ireland\\
  Shanghai Jiao Tong University, Shanghai, China\\
  \texttt{chksomyajit@sjtu.edu.cn}
  \And
  Angshuman Jana\\
  Indian Institute of Information Technology, Guwahati, India\\
  \texttt{angshuman@iiitg.ac.in}
  \And
  Avijit Gayen\\
  Indian Institute of Information Technology, Guwahati, India\\
  Techno India University, West Bengal, Kolkata, India\\
  \texttt{avijit.gayen@iiitg.ac.in, avijit.g@technoindiaeducation.com}
}

\renewcommand{\shorttitle}{Bridging Through Absence}

\hypersetup{
  pdftitle={Bridging Through Absence: How Comeback Researchers Bridge Knowledge Gaps Through Structural Re-emergence},
  pdfsubject={cs.SI, physics.soc-ph},
  pdfauthor={Somyajit Chakraborty, Angshuman Jana, Avijit Gayen},
  pdfkeywords={comeback researchers, citation networks, knowledge bridging, early-career researchers, bibliometrics, community silos, research discontinuity, machine learning},
}

\begin{document}
\maketitle

\begin{abstract}
Understanding the role of researchers who return to academia after prolonged inactivity---termed ``comeback researchers''---is crucial for developing inclusive models of scientific careers. This study investigates the structural and semantic behaviors of comeback researchers, focusing on their role in cross-disciplinary knowledge transfer and network reintegration. Using the AMiner citation dataset, we analyze 113{,}637 early-career researchers and identify 1{,}425 comeback cases based on a \emph{three-year-or-longer} publication gap followed by renewed activity. We find that comeback researchers cite 126\% more distinct communities and exhibit 7.6\% higher bridging scores compared to dropouts. They also demonstrate 74\% higher gap entropy, reflecting more irregular yet strategically impactful publication trajectories. Predictive models trained on these bridging- and entropy-based features achieve a 97\% ROC-AUC---far outperforming the 54\% ROC-AUC of baseline models using traditional metrics like publication count and h-index. Finally, we substantiate these results via a multi-lens validation. These findings highlight the unique contributions of comeback researchers and offer data-driven tools for their early identification and institutional support.
\end{abstract}


\section{Introduction}
\label{sec:intro}
Scientific progress is not solely the product of individual ideas but emerges from the evolving structures that connect them in citation flows, collaboration ties, and disciplinary transitions. These interconnections shape the trajectory of research fields and the position of individual scholars within the broader scientific ecosystem. Over the past two decades, the integration of network science and bibliometrics has provided robust methodological frameworks for modelling such structures. Citation networks, in particular, represent scholarly communication as graphs, where papers are nodes, and citations form directional edges that signify intellectual influence and epistemic lineage. On the other hand, collaboration network, can be represented by co-authorship relationship among the researchers, where authors are represented as nodes and their collaborative relationship as edge in the network.  These networks are dynamic, reflecting both the emergence of new research themes and the consolidation or decline of older paradigms. Within these evolving systems, certain structural roles—such as hubs, authorities, and particularly bridges—have been recognized as critical to the flow of knowledge across disciplinary and conceptual boundaries.

\par Amid these structural dynamics, early-career researchers (ECRs) play a pivotal role in shaping the future contours of scientific inquiry. Typically defined as researchers under the age of 35, often pursuing or having recently completed a doctoral degree~\cite{nicholas2017early,nicholas2019so,gayen2023characterization,gayen2024can,10307427}, ECRs are characterized by high intellectual curiosity, adaptability, and the willingness to engage in emerging or interdisciplinary fields. Despite their limited academic tenure, ECRs have been shown to contribute significantly to innovative research, especially when embedded in collaborative environments that offer access to mentorship, resources, and diverse scholarly perspectives~\cite{evans2018introduction}. Their contributions are not merely incremental; they often bring new methods, disruptive questions, and transdisciplinary perspectives that challenge established norms.

\par However, the early-career phase is also marked by a heightened risk of professional discontinuity. Numerous studies have highlighted the precarious position of ECRs, citing structural challenges such as short-term contracts, funding scarcity, pressure to publish in high-impact venues, and difficulties balancing academic work with personal or family responsibilities~\cite{Stratford2024} \cite{Krauss2023}. These pressures frequently result in prolonged gaps in publishing or complete disengagement from academic research. This phenomenon is commonly referred to as academic dropout as explored in our previous study~\cite{gayen2025comeback}. While dropout is often framed as a terminal stage in academic careers, such a binary perspective overlooks a critical but underexamined phenomenon. A small subset of researchers return to academic publishing after extended periods of inactivity—these are the individuals we term ``comeback researchers''. A recent bibliometric study by Sidelmann and Grimstrup~\cite{sidelmann2025competition} further evidences this trend, showing that only 50\% or fewer researchers remained active a decade after their initial publication in the 2010s, marking a sharp rise in academic attrition compared to earlier cohorts. These individuals, whom we refer to as ``comeback researchers''~\cite{gayen2025comeback} offer a unique lens through which to examine the non-linear dynamics of scientific careers and the structural reintegration of dormant expertise.

\par Despite their potential significance, comeback researchers remain largely invisible in mainstream science-of-science analyses. Most prior research has focused either on those who sustain continuous productivity or those who drop out entirely, leaving a blind spot in our understanding of academic re-entry and re-integration. This gap is particularly striking given the potential structural distinctiveness of comeback researchers. The structural and intellectual reintegration of comeback researchers opens up several intriguing possibilities and also open up various questions---{\em Do they resume former research lines, or does their re-entry reflect a deeper transformation—perhaps shaped by industry experience, exposure to new domains, or personal reflection?} Their temporal discontinuity may function as a crucible, refining scholarly focus and enabling more eclectic, interdisciplinary engagements upon return. As such, comeback researchers could serve as latent bridges across research silos—facilitating knowledge diffusion in ways distinct from both active and dropout peers. It is unclear whether these individuals simply resume previous research trajectories or whether their return coincides with a broader reconfiguration of their academic roles. For instance, having experienced time away from academia, comeback researchers may bring with them exposure to new disciplines, industry insights, or non-traditional knowledge sources, allowing them to serve as boundary-crossers within the citation network. Alternatively, the disruption itself may reorient their publishing strategies toward more diverse or interdisciplinary collaborations, enhancing their bridging capacity within the scientific landscape. To address these questions, we frame our study around three core research questions:
\begin{itemize}
    \item \textbf{RQ1}: \textit{To what extent do comeback researchers engage in cross-community knowledge transfer, as compared to their dropout counterparts, based on citation and community interaction patterns?}
    \item \textbf{RQ2}: \textit{Do comeback researchers exhibit a higher tendency to structurally bridge knowledge silos within scientific citation networks?}
    \item \textbf{RQ3}: \textit{Can network-based bibliographic metrics reliably distinguish between researchers who permanently drop out and those who re-emerge after discontinuation?}
\end{itemize}

\par
\begin{figure}[t]
  \centering
  \includegraphics[width=\textwidth]{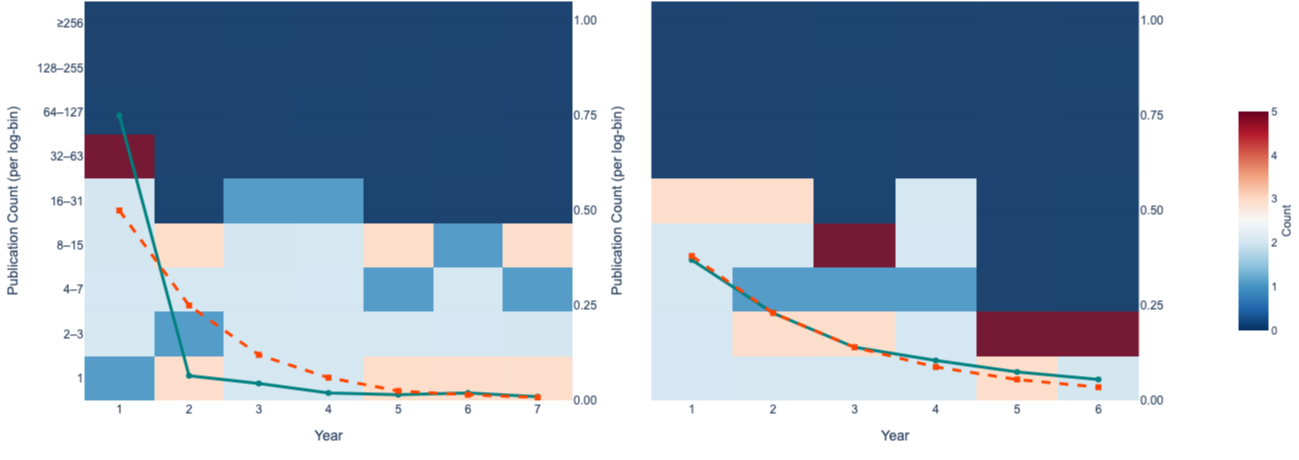}
  \caption{Figure shows the publication-count heatmaps (log-bins) for (Left) Dropout and (Right) Comeback authors. The heatmaps are overlaid with scaled probability curves, where the empirical probability is shown as a solid line and the model fit as a dashed line. The x-axis represents the year index, and the y-axis represents the logarithmic bin of publication counts.}
  \label{fig:heatmaps}
\end{figure}

\begin{figure}[t]
  \centering
  % Use the vector PDF for best print quality; switch to .png if you prefer
  \includegraphics[width=\linewidth]{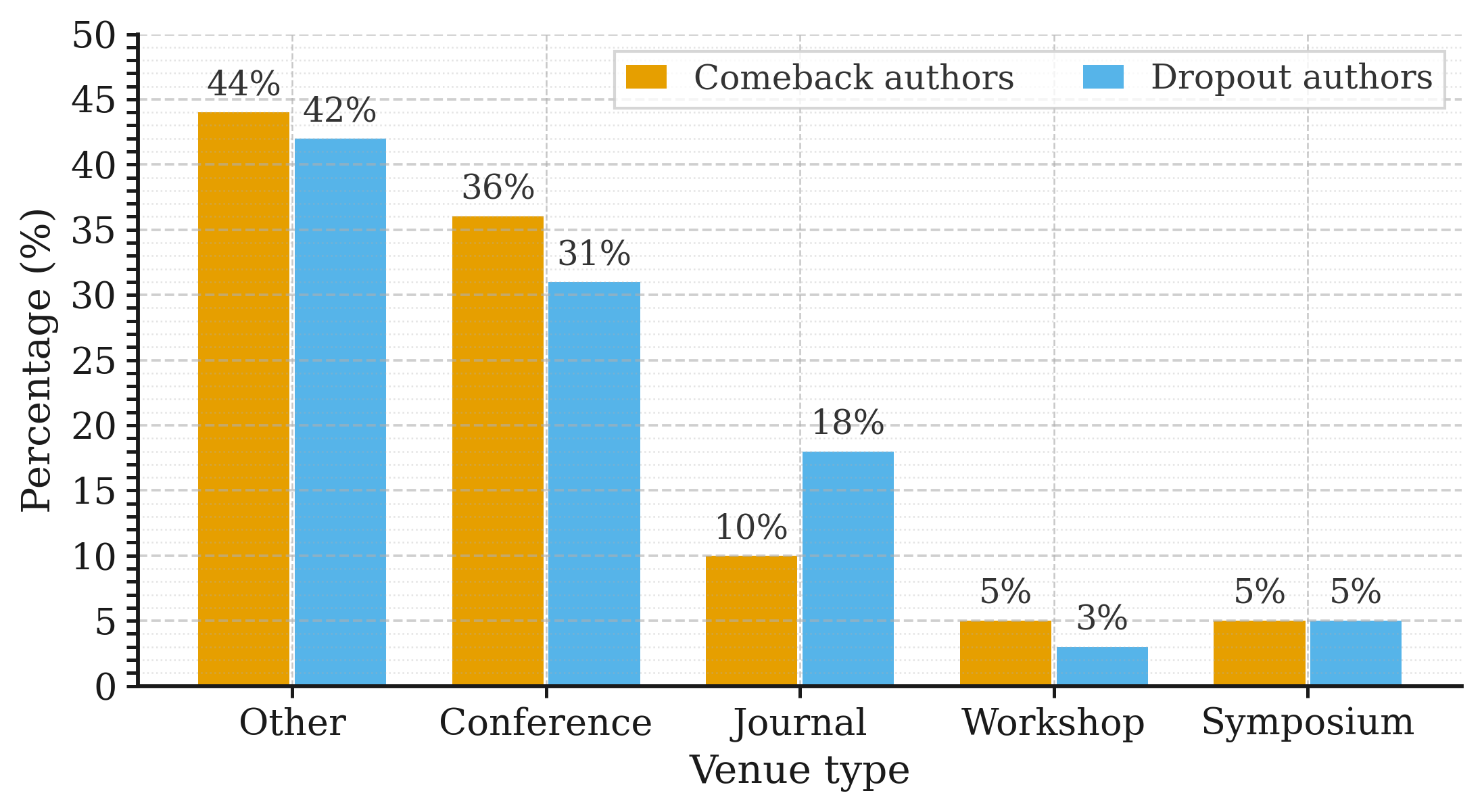}
  \caption{Figure shows the grouped bar chart showing the percentage share of different venue types for Comeback versus Dropout authors. The y-axis shows the percentage (\%), and the x-axis categorizes the venue types: Other, Conference, Journal, Workshop, and Symposium.}
  \label{fig:venue_grouped_bars}
\end{figure}

\begin{figure}[t]
  \centering
  \includegraphics[width=1.01\linewidth,height=0.55\linewidth]{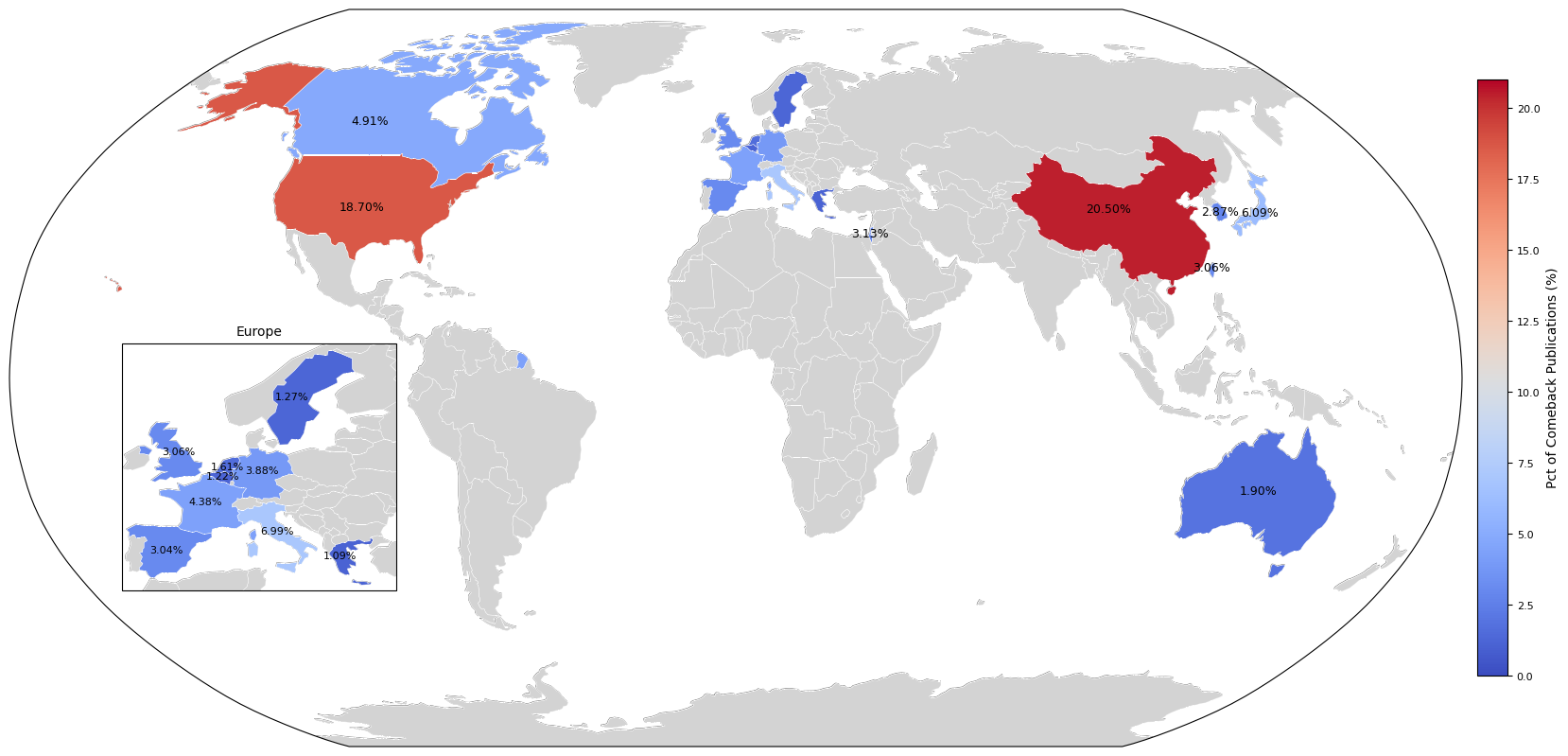}
  \caption{The figure represents global geographic distribution of comeback-publication percentages by country. A zoomed inset focuses on Europe. The color intensity or value on the map corresponds to the percentage of comeback publications for each country.}
  \label{fig:worldmap}
\end{figure}

 To see why comeback researchers deserve special attention, we first present three empirical snapshots drawn from our AMiner data (Figures~\ref{fig:heatmaps}–\ref{fig:worldmap}). Figure~\ref{fig:heatmaps} compares dropout and comeback authors in terms of their publication‐count distributions (binned logarithmically) alongside scaled probability curves. On the left, the dropout cohort’s heatmap (Years 1–7) is overlaid with its empirical and model dropout‐probability curves. We fit an exponential–decay model to the year index $t$, $p_{\text{model}}(t)=p_0 e^{-\lambda t}$ (curves min–max scaled for visualization). In Fig.~\ref{fig:heatmaps}, the empirical probability is the solid line and the model fit is the dashed line; on the right, the comeback cohort’s heatmap (Years 1–6) is similarly overlaid with its empirical and model comeback‐probability curves. Although both sets of probability curves decline over time, the comeback authors (right) concentrate their publications in different log‐bins compared to permanent dropouts (left), suggesting a distinct distribution of productivity upon re‐entry.  

\par In figure~\ref{fig:venue_grouped_bars}, we observe the differences in venue‐type shares for both cohorts. For the comeback venues: “Other” accounts for 44 \%, “Conference” 36 \%, “Journal” 10 \%, “Workshop” 5 \%, and “Symposium” 5 \%. By contrast, for the dropout cohort: “Other” 42 \%, “Conference” 31 \%, “Journal” 18 \%, “Symposium” 5 \%, and “Workshop” 3 \%. The noticeably larger “Journal” slice among dropouts (18 \%) versus comeback authors (10 \%) and the larger “Other” slice among comeback authors (44 \%) highlight how comeback researchers publish in a broader mix of venues. 
\par
% \textcolor{blue}{In Figure~\ref{fig:venue_grouped_bars}, we observe clear differences in venue mixes between cohorts. For comebacks: “Other” accounts for 44\,\%, “Conference” 36\,\%, “Journal” 10\,\%, “Workshop” 5\,\%, and “Symposium” 5\,\%. For dropouts: “Other” 42\,\%, “Conference” 31\,\%, “Journal” 18\,\%, “Symposium” 5\,\%, and “Workshop” 3\,\%.}

The higher \emph{Conference\,+\,Other} share among comeback authors (80\,\% vs.\ 73\,\%) together with their lower journal share (10\,\% vs.\ 18\,\%) is consistent with a \textbf{re-entry strategy that favors fast-cycle, boundary-crossing outlets}. Conferences and “Other” (e.g., preprints/tech reports, book chapters, demos) typically host earlier-stage, applied, or cross-domain work and draw heterogeneous audiences. In contrast, the relatively larger journal share among dropouts suggests earlier specialization that did not diversify before exit. If comeback authors disproportionately use outlets that sit at \emph{interfaces} between topical communities, then their citation ego-nets should span \emph{more communities} and occupy \emph{more connective} positions. This venue-level signal motivates our network-level tests. More Conference/Other at (re-)entry leads to more cross-community citations. A higher Journal share corresponds to lower cross-community engagement and lower bridging.
% \textcolor{red}{What is the inference of this observation? How is it become the motivation of our work?-- Not clear.} 

\par Finally, Figure~\ref{fig:worldmap} displays the global geographic distribution of comeback‐publication percentages by country (with a zoom‐inset on Europe). Here, China (20.5 \%) and the United States (18.7 \%) lead in comeback shares, whereas most European nations re‐enter at much lower percentages (e.g., Germany 3.9 \%, France 4.4 \%, Sweden 1.3 \%). This worldwide pattern underscores that the phenomenon of returning to academic publishing is particularly pronounced in a handful of countries, further motivating our focus on the structural and contextual factors that differentiate comeback researchers from both dropouts and consistently active peers. Guided by these empirical differences in venue mix and geography, we study whether comeback researchers systematically \emph{bridge} knowledge silos—\emph{and if so, why}—by examining their positions in dynamic citation networks and their diversity of cross-community engagement.


\par The objective of this study is to systematically characterize comeback researchers through structural, semantic, and temporal dimensions. Guided by the three research questions, we aim to (i) measure their engagement in cross-community citation flows, (ii) evaluate their bridging roles within dynamic citation networks, and (iii) build predictive models that distinguish comeback authors from permanent dropouts. Our broader goal is to uncover whether career discontinuity, far from signaling decline, may in fact reposition certain researchers as pivotal agents in interdisciplinary knowledge transfer. This study posits that the structural trajectories of comeback researchers—marked by disruption followed by re-entry—may predispose them to occupy unique positions in citation networks, particularly as brokers or bridges between otherwise disconnected communities. While hubs typically accumulate influence through sustained output and citation accrual, bridges enable the diffusion of knowledge across disciplinary or thematic silos, facilitating innovation through recombination. Identifying whether comeback researchers disproportionately assume these bridging roles is of both theoretical and practical interest. Theoretically, it challenges cumulative advantage models that prioritize continuous productivity. Practically, it informs institutional policies around reintegration, mentorship, and support for researchers with non-linear careers. In addition to their potential structural distinctiveness within citation networks, comeback researchers may also exhibit unique semantic and temporal behaviors that warrant systematic investigation. Their re-entry into academia is often characterized by irregular publication rhythms, thematic shifts, or broader topical engagement, reflecting either a reorientation of scholarly focus or the integration of new perspectives acquired during their hiatus. Temporally, their return may not follow a gradual reintegration pattern but could involve concentrated bursts of activity indicative of renewed engagement or strategic publishing efforts. Moreover, modeling these attributes enables the development of predictive frameworks that differentiate between researchers likely to re-engage with academic publishing and those who permanently disengage. Such insights not only advance theoretical understanding of non-linear academic careers but also offer practical tools for early identification and institutional support.

\par To answer these questions, we propose a comprehensive empirical framework grounded in dynamic citation network analysis and career trajectory modeling. Using the AMiner academic graph dataset, we first classify authors into three cohorts—comeback, dropout, and consistently active—based on temporal patterns of publication continuity and discontinuity. We then construct yearly citation networks to capture the evolving structure of scholarly communication over time. To identify topical groupings within these networks, we apply community detection using the Louvain algorithm. This helps to identify groups of papers that cite one another more densely than the rest of the graph. This step serves two purposes. First, it gives a data-driven notion of “field” or “topic cluster” without relying on venue tags. Second, it lets us measure how often an author engages across these groups versus staying within one. This enables us to analyze whether comeback researchers cite across, or are cited by, multiple research communities—an indicator of interdisciplinary influence. Using these graphs and communities, we compute a small set of indicators that summarize (i) where an author sits in the network, (ii) how broadly they engage across groups, and (iii) how regular or irregular their career timing is. We compare comeback and dropout cohorts on these indicators and assess whether the observed differences are statistically meaningful. Predicting comeback behavior is useful for policy and practice. Early identification enables targeted support (e.g., re-entry fellowships, bridge funding, mentorship) for scholars with non-linear careers. For funders and institutions, it offers a fairer alternative to productivity-only screening, which can miss latent integrators who are valuable for cross-domain knowledge flow. Statistical comparisons are performed to evaluate whether comeback researchers differ significantly from their dropout counterparts across these dimensions. Furthermore, we develop interpretable machine learning models to test whether comeback behavior is predictable using these features, thereby offering a data-driven foundation for early identification and support of re-entry trajectories. The major novel contributions of this work can be listed as below:
\paragraph{Our Contributions:}
\begin{itemize}
  \item We provide the first large-scale, cohort-level characterization of \emph{comeback researchers} defined by a \textbf{$\leq 3-year$} publication gap followed by renewed activity.
  \item We show that comebacks engage more broadly across detected communities and occupy more connective positions than dropouts, evidencing a \emph{bridging} role in citation networks.
  \item We introduce simple, interpretable indicators that encode cross-community engagement and temporal irregularity, and demonstrate that they \emph{predict} comebacks with high ROC–AUC.
  \item We release a clear, reproducible pipeline for dynamic network construction, cohort labeling, and evaluation, suitable for extension to other fields and time windows.
  \item We discuss policy implications for re-entry support, arguing for evaluation criteria that go beyond productivity-only metrics.
\end{itemize}

\par The rest of the paper is organized as follows: Section~\ref{s:related} reviews related work to figure out the scope of our work. In section~\ref{sec:metrics}, we  define the bibliographic metrics used in our analysis. A detailed description of our data preprocessing, author classification, citation network construction, metric formulation, and modeling strategy is provided in Section~\ref{sec:methodology}, with each step designed to progressively uncover the distinct roles of comeback researchers.   Section~\ref{sec:methodology} describes the methodology adopted in our work which include dataset description, cohort labeling, network construction, community detection, and modeling pipeline. In  section~\ref{sec:results}, we present empirical results and the discussion of the implications is integrated there. Finally, we conclude  in section~\ref{sec:conclusion} where we also discuss the limitation of the work and possible future directions.

\section{Related Works}\label{s:related}
This section reviews prior studies relevant to early-career researchers (ECRs), research discontinuation, and structural knowledge integration in citation networks. We organize the discussion into five key strands: trends in scientific collaboration, challenges of early career researchers, discontinuation and return patterns, structural bridging in citation networks, and predictive modeling of academic trajectories.

\paragraph{\bf Trends in Scientific Collaboration}
The need for collaborative research work has long been recognized, particularly for addressing complex, interdisciplinary problems. Over the past few decades, the volume and diversity of scientific collaboration have grown substantially~\cite{borgman2002scholarly,wagner2015continuing}. Advances in communication technology have facilitated both local and international collaborations~\cite{kong2016exploiting1,abbasi2010evaluating}. Several studies have emphasized the rise in international scientific collaborations~\cite{leclerc1994international} and the increasing demand for researchers capable of bridging disciplinary divides~\cite{montoya2018fast}. Collaborative efforts are shown to increase knowledge exchange, with many researchers identifying “increased knowledge” as the most valuable outcome~\cite{melin2000pragmatism}.

\paragraph{\bf Early-Career Researchers: Roles and Challenges}
ECRs—commonly defined as Ph.D. candidates or postdoctoral researchers—play a vital role in advancing scientific innovation~\cite{nicholas2019so,bridle2013preparing,laudel2019field,wang2019early,allen2019open,morriss2019enhancing,gayen2023characterization,thomsen2021introduction}. Their efforts often drive interdisciplinary research~\cite{bridle2013preparing} and benefit from international exposure~\cite{djerasimovic2020constructing}. However, they face several challenges, including time management, funding scarcity, and limited institutional recognition~\cite{ortlieb2018makes,maer2019skill,barkhuizen2021identity}. In interdisciplinary settings, these barriers are exacerbated by steep learning curves, publication pressure, and limited supervisory support~\cite{andrews2020supporting,horta2018phd,castello2021perspectives}. Additionally, ECRs often encounter systemic instability, described as a `risk-career' environment in higher education policy~\cite{skakni2019hanging,castello2015researcher}. Exploitation in peer review~\cite{mcdowell2019co} and challenges in publishing in high-impact venues~\cite{drosou2020overcoming} further compound these issues.

\paragraph{\bf Research Discontinuation and Comeback Patterns}Despite increasing attention to research careers, the phenomenon of discontinuation remains underexplored. Prior works have observed that publication gaps—often due to life events, job changes, or burnout—are particularly common among ECRs~\cite{jadidi2018gender,larcombe2022makes,krauskopf2018analysis}. Gender-specific barriers, such as unequal access to funding and collaborations, are also linked to higher dropout rates among women~\cite{jadidi2018gender}. In our recent study~\cite{gayen2025comeback}, we analysed over 113,000 researchers and found that nearly 90\% of those with early-career publication gaps never returned to academia. Yet a small fraction, termed \emph{comeback researchers}, do resume publication after several years, often re-entering through interdisciplinary or translational domains. These individuals, while rare, exhibit distinct patterns of knowledge integration and often bridge disconnected areas of scientific inquiry.

\paragraph{\bf Citation Network Structure and Knowledge Bridging}Scientific progress is deeply influenced by the structure of citation and co-authorship networks. Nodes with high \emph{betweenness centrality} are often brokers across research communities, enabling novel recombinations of knowledge~\cite{chen2025approach}. Several studies have shown that these bridge-like positions—also known as spanning structural holes—are correlated with higher citation impact~\cite{Fleming2007Brokerage}, \cite{Uzzi2013Atypical}, \cite{Wang2015Interdisciplinarity}, \cite{Wang2023StructuralHolesPhysics}. Citation network studies have emphasized the role of interdisciplinary connections in fostering innovation and knowledge diffusion~\cite{Wang2015Interdisciplinarity},\cite{Wang2023StructuralHolesPhysics}. However, many research areas remain isolated as \emph{knowledge silos}, with limited citation flows between them~\cite{cunningham2025knowledge}. By applying dynamic community detection, recent work in explainable AI has revealed significant “knowledge gaps” between foundational and contemporary research areas, underscoring the need for agents that actively connect these communities~\cite{cunningham2025knowledge}. Comeback researchers—returning after career interruptions—may fill this structural role by importing diverse knowledge gained during hiatuses.

\paragraph{\bf Predictive Modeling of Research Careers}
The emergence of large-scale scholarly databases has enabled predictive modeling of research trajectories. Several studies have shown that early-career publication counts and network features (e.g., co-author centrality) are strong predictors of long-term impact \cite{Li2019NatComm, Momeni2023EPJDS}. In our recent studies \cite{gayen2024can, gayen2025comeback} we also see the importance of such metrics as predictors of impact. It is seen that the inclusion of network centrality measures significantly improved the prediction of academic promotions~\cite{wapman2022quantifying, li2022untangling}. However, recent work emphasizes the importance of randomness and serendipity in scientific careers. The \emph{random impact rule} suggests that major scientific breakthroughs can occur at any point in a career~\cite{sinatra2016quantifying, wang2019early}. Sinatra et al.~\cite{sinatra2016quantifying} in 2016 introduced the concept of a person-specific ``Q-factor'' that reflects inherent impact potential, independent of age or publication order. These models provide a framework for identifying researchers with comeback potential—even those with non-linear or erratic publication histories. The observed “hot streak” phenomenon further supports the possibility of late-career resurgence~\cite{liu2018hot,oliveira2023hot}.

\par While the challenges of ECRs and dropout phenomena have been partially explored, the structural contributions and reintegration potential of comeback researchers remain underexamined. Most existing models and policies assume linear, uninterrupted careers. Our study fills this gap by empirically analyzing comeback researchers from a network-theoretic perspective—evaluating their roles in bridging communities and enhancing cross-domain knowledge transfer. In doing so, we aim to shift focus from mere attrition metrics to the latent value of re-engagement and structural reinvention within science.

\section{Bibliographic Metrics}
\label{sec:metrics}This section details the quantitative indicators used to characterize and distinguish comeback researchers. The rationale for this section is twofold: first, to formally define the established and novel metrics that operationalize our investigation into cross-community bridging and temporal career patterns; and second, to provide a transparent foundation for our analytical and predictive modeling pipeline. We begin by establishing the foundational concepts of the citation network and research communities, which are central to our metric definitions. We then systematically present the established bibliographic metrics that serve as our baseline, followed by our three novel proposed metrics designed to capture bridging behavior and career irregularity. Finally, we subject these proposed metrics to a multi-faceted validation framework to ensure their robustness, validity, and fitness for purpose.

\paragraph{Foundational Concepts: Citation Network and Communities}Our analysis is grounded in a dynamic, author-centric view of the scientific citation network. The foundational elements are defined as follows:

\par{\bf Citation Network:} We construct a directed graph \( G_t = (V_t, E_t) \) for each year \( t \), where \( V_t \) is the set of all published papers up to year \( t \), and a directed edge \( (p_i \to p_j) \in E_t \) exists if paper \( p_i \) cites paper \( p_j \).
\par{\bf Research Communities:} To identify topical groupings in a data-driven manner, we apply the Louvain algorithm to the paper-paper citation network for each year. This yields a partition of papers into communities, where papers within a community cite each other more densely than they cite papers outside it. Let \( c(p) \) denote the community assignment of a paper \( p \). This concept of a ``community'' serves as our operational definition of a research field or topic silo.
\par{\bf Observation Window:} For a fair comparison, all author-level metrics are computed within a non-leaky observation window. For a comeback author \( a \), this window spans from their first publication year up to the year immediately preceding their return. For a matched dropout author, the window is of equal length, ending at their last active year.

All subsequent metrics are computed based on an author's activity within this framework. Let \( P_a \) denote the set of papers authored by \( a \) within the observation window, and \( R(p) \) the set of references of a paper \( p \in P_a \).

\subsection{Established Metrics} In this section, we formally define several existing well-known metrics in the context of citation network used in our work. \paragraph{\textit{Publication Count(P):}} The total number of publications authored by a researcher is given by: \begin{equation} P = \sum_{i=1}^{T} P_{i} \end{equation} where: \begin{itemize} \item \( P_{i} \) is the number of publications in {\em i}-th year by the author. \end{itemize} \paragraph{\textit{Citation Count({\em C}):}}Citation count calculates the frequency with which other researchers have referenced an author's work in their papers. To calculate it, we consider all citations received from each of his or her publications cumulatively during his research tenure. This is a widely used, straightforward metric to assess how much a researcher's work has affected and influenced their field. \begin{equation} C = \sum_{i=1}^{P} C_i \end{equation} where: \begin{itemize} \item \( C_i \) is the quantity of citations that the \( i \)th publication has received. \item \( P \) is the amount of publications overall. \end{itemize} \paragraph{\textit{h-index ({h}):}} The {\em h}-index~\cite{hindex2005} is a widely used measure to estimate the impact and productivity of a researcher’s publications. It is defined as the maximum value of \( h \) such that the researcher has at least \( h \) papers, each of which has been cited at least \( h \) times. \begin{equation} h = \max \{ h \mid \text{at least } h \text{ papers have } \geq h \text{ citations} \} \end{equation}

\paragraph{Cross-Community Citation Rate (XCC).}This metric evaluates interdisciplinary citation behavior. For an author $a$, let $P_a$ be their set of papers and $R(p)$ be the references of a paper $p$. Let $c(p)$ be the community of $p$ (from Louvain detection~\cite{de2011generalized}). The XCC is: \begin{equation} \text{XCC}(a) = \frac{1}{|P_a|} \sum_{p \in P_a} \frac{|\{q \in R(p) : c(q) \neq c(p)\}|}{|R(p)|} \end{equation} This value captures the extent to which an author cites outside their own disciplinary silo, making it useful for detecting knowledge diffusion across communities \cite{cunningham2025knowledge}. Note that $B$ and XCC differ only by weighting (edges vs.\ papers) and therefore track similar behavior.

\subsection{Proposed Metrics}
\label{sec:proposed_metrics}

While established bibliometric indicators capture productivity and impact, they are often ill-suited for characterizing the unique structural and temporal patterns of researchers with non-linear careers. To specifically quantify the behaviors we hypothesize for comeback researchers---namely, their role as knowledge bridges and their irregular career trajectories---we propose three novel metrics. These are designed to measure an author's \textbf{cross-community engagement} (\textit{Bridging Score}), the \textbf{breadth of their own research} (\textit{Authored-Community Count}), and the \textbf{temporal irregularity} of their publication history (\textit{Gap Entropy}). The rationale for these metrics is to move beyond volume-based assessment and instead capture the nuanced ways in which researchers, especially those re-integrating after a hiatus, connect disparate parts of the scientific landscape and manage their scholarly output over time.

\noindent\textbf{\em Bridging Score (B)}

The Bridging Score is an edge-weighted metric that quantifies the extent to which an author's citations connect different research communities. It is defined as the fraction of an author's outgoing citation edges that point to papers outside the citing paper's own community. Formally, for an author \(a\), let \(E_a\) be the multiset of all outgoing citation edges from their papers (i.e., each \(p \to q\) where \(p\) is a paper by \(a\) and \(q\) is a reference of \(p\)). The Bridging Score is calculated as:

\begin{equation}
B(a) = \frac{|\{(p \to q) \in E_a : c(q) \neq c(p)\}|}{|E_a|}    
\end{equation}

Here, \(c(p)\) and \(c(q)\) denote the communities of paper \(p\) and reference \(q\), respectively. A higher \(B(a)\) indicates a greater propensity to import knowledge from outside an author's immediate research silo, suggesting a stronger bridging role in the citation network.

\par While both XCC and $B$ measure cross-community engagement, $B$ is edge-weighted (normalizing by the total number of citation edges) and captures overall cross-silo \emph{intensity}, whereas XCC averages the cross-community share \emph{per paper}, highlighting how evenly that engagement is distributed across an author's publication portfolio.

\vspace{0.5cm}
\noindent\textbf{\em Distinct Authored-Community Count (ACC)}

This metric captures the topical breadth of an author's own body of work by simply counting the number of distinct research communities in which they have published. It is defined as the cardinality of the set of communities represented by all papers authored by \(a\) within the observation window:

\begin{equation}
\text{ACC}(a) = \left|\{c(p) : p \in P_a\}\right|  
\end{equation}

where \(P_a\) is the set of papers authored by \(a\). A higher ACC value signifies that the author's research spans a wider range of topics or fields, as defined by the community structure of the citation network.

\vspace{0.5cm}
\noindent\textbf{\em Gap Entropy (\(H_g\))}

Gap Entropy measures the irregularity or ``burstiness'' of an author's publication timeline, which is a hypothesized characteristic of non-linear careers. It is calculated using the information entropy of the distribution of gaps (in years) between consecutive publications. For an author \(a\), let \(\Delta_i = y_{i+1} - y_i\) be the sequence of inter-publication gaps derived from their sorted publication years. We first compute the probability distribution \(q_j\) over these gap lengths (e.g., from a histogram). The Gap Entropy is then defined as:

\begin{equation}
  H_g(a) = -\sum_{j} q_j \log q_j  
\end{equation}

A perfectly regular publisher, who publishes every year, would have a Gap Entropy of 0. Conversely, a higher \(H_g\) indicates a more unpredictable and erratic publication pattern, with a mix of short and long gaps. All $H_g$ values use $\log_2$ and gap-length histograms with a bin width of 1 year.

\begin{table}[ht]
\centering
\resizebox{\textwidth}{!}{%
\begin{tabular}{|p{4cm}|p{3cm}|p{6.5cm}|}
\hline
\textbf{Existing Metric} & \textbf{Reference / Origin} & \textbf{Use in This Study} \\
\hline
Publication Count ($P$) & Standard bibliometric & Baseline for author productivity (volume control) \\
Citation Count ($C$) & Standard bibliometric & Baseline for impact (volume control) \\
h-index ($h$) & Hirsch et al.~\cite{hindex2005} & Career performance baseline; used in baseline-only models \\
Cross-Community Citation Rate (XCC) & Cunningham (2025)\cite{cunningham2025knowledge} & Paper-weighted share of citations outside the paper’s community; reported in Fig.~\ref{fig:crosscomm} \\
\hline
\textbf{Proposed Metric} & \textbf{Reference / Origin} & \textbf{Use in This Study} \\
\hline
Bridging Score ($B$) & This work & Edge-weighted fraction of an author’s outgoing citations that cross communities; main structural indicator (Figs.~\ref{fig:communitybox}, \ref{fig:score_ci}, \ref{fig:heatmap}, \ref{fig:window}, \ref{fig:survival}) \\
Distinct Authored-Community Count (ACC) & This work & Number of distinct communities spanned by the author’s own papers; breadth of topical engagement (Figs.~\ref{fig:communitybox}, \ref{fig:score_ci}) \\
Gap Entropy ($H_g$) & This work & Temporal irregularity of publishing (career burstiness); used in Fig.~\ref{fig:gapentropy} \\
\hline
\end{tabular}%
}
\caption{Summary of bibliographic metrics used in this study, categorized as Existing and Proposed. For each metric, the table lists its name, reference or origin, and its specific use within this study.}
\label{tab:metrics_summary}
\end{table}

\begin{figure}[ht]
\centering
\includegraphics[width=0.9\linewidth]{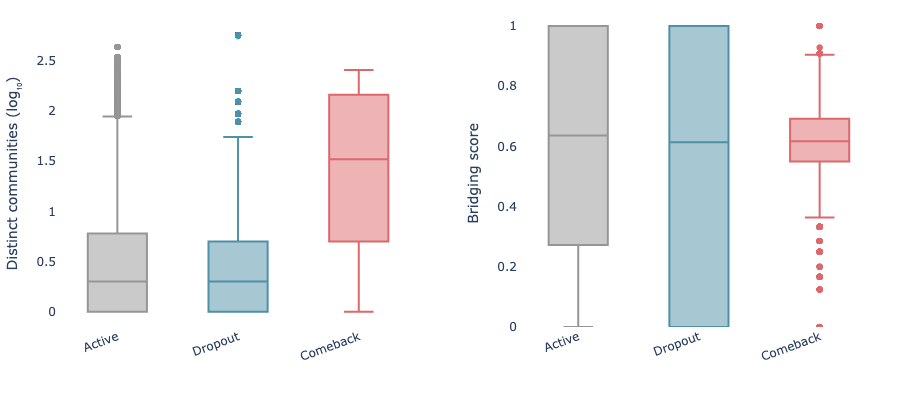}
\caption{Figure represents distribution of Distinct Communities (log-scaled, left) and Bridging Score (right) across author categories (Comeback, Dropout, Active). The y-axis for the left plot is the log-scaled count of distinct communities. The y-axis for the right plot is the Bridging Score value.}
\label{fig:communitybox}
\end{figure}

\begin{figure}[ht]
\centering
\includegraphics[width=0.95\linewidth]{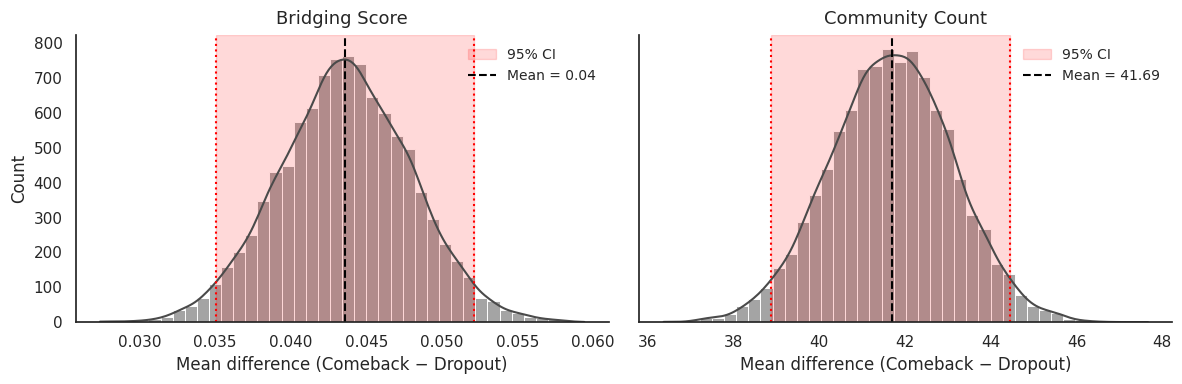}
\caption{Figure shows the  bootstrapped distributions of the mean difference (Comeback – Dropout) for (left) Bridging Score and (right) Distinct Authored-Community Count (ACC). The x-axis shows the mean difference value, and the y-axis shows the frequency from bootstrapping. The vertical dashed lines indicate the observed mean differences.}
\label{fig:score_ci}
\end{figure}

\subsection{Metric Validation}
\label{sec:metric_validation}

To ensure the proposed metrics—Bridging Score (\(B\)), Distinct Authored-Community Count (ACC), and Gap Entropy (\(H_g\))—are robust, meaningful, and fit for purpose, we subject them to a multi-faceted validation framework. This section details the rationale and results of four key validation tests: statistical sanity, correlational analysis, discriminant validation, and robustness checks.

\subsubsection{Statistical Sanity Check}
\label{sec:sanity_check}

\textbf{Rationale:} A fundamental requirement for any metric is that it behaves in an intuitive and mathematically sound manner. The sanity check verifies that the metrics move monotonically in the expected direction as the underlying behavior they are designed to capture becomes more pronounced.

\textbf{Observation:} Our analysis confirms the expected monotonic behavior for all proposed metrics:
\begin{itemize}
    \item \textbf{Bridging Score (\(B\)) and XCC:} Both metrics increase consistently as a larger share of an author's references cross community boundaries. \(B\) provides an edge-weighted perspective, while XCC offers a paper-weighted one, but both track the same core behavior of cross-community citation.
    \item \textbf{Authored-Community Count (ACC):} This count increases directly as an author's own publications span a greater number of distinct research communities, accurately reflecting topical breadth.
    \item \textbf{Gap Entropy (\(H_g\)):} For an author with a perfectly regular annual publication record (all gaps \(\Delta_i = 1\)), the gap entropy is zero. The value of \(H_g\) grows as the probability mass shifts towards longer or more varied inter-publication gaps, correctly quantifying temporal irregularity.
\end{itemize}

\textbf{Inference:} The proposed metrics pass the basic sanity check, confirming their mathematical formulation correctly translates the intended scholarly behaviors—bridging, breadth, and irregularity—into quantifiable scores.

\begin{figure}[ht]
\centering
\includegraphics[width=0.7\linewidth]{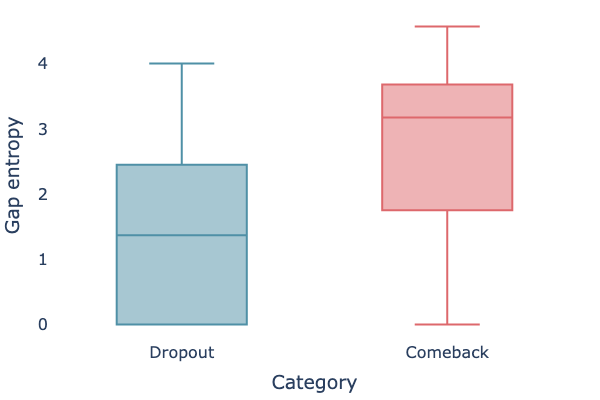}
\caption{Figure represents the distribution of Gap Entropy for Comeback vs. Dropout authors. The y-axis represents the density or frequency, and the x-axis represents the Gap Entropy value.}
\label{fig:gapentropy}
\end{figure}

\begin{figure}[ht]
\centering
\includegraphics[width=0.6\linewidth]{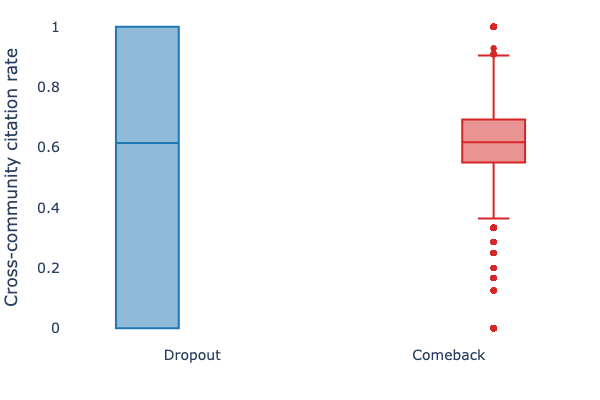}
\caption{Figure shows the boxplot of Cross-Community Citation Rate (XCC) for Comeback vs. Dropout authors. The y-axis shows the Cross-Community Citation Rate.}
\label{fig:crosscomm}
\end{figure}

\subsubsection{Correlational Validation with Known Metrics}
\label{sec:correlational_validation}

\textbf{Rationale:} To establish convergent and divergent validity, we examine the relationships between the new metrics and established bibliometric indicators. We expect the new metrics to capture unique dimensions of a research career not reflected by traditional volume-based measures.

\textbf{Observation:} The correlation analysis reveals a clear pattern:
\begin{itemize}
    \item \textbf{Convergent Validity:} The edge-weighted Bridging Score (\(B\)) and the paper-weighted Cross-Community Citation Rate (XCC) are strongly positively associated (Spearman \(\rho \approx 0.75\)), as expected, since they are two operationalizations of the same cross-community engagement construct.
    \item \textbf{Divergent Validity:} Crucially, all three cross-community indicators (\(B\), XCC, and ACC) show near-zero correlations with volume metrics such as total publication count (\(P\)), total citation count (\(C\)), and the h-index (\(h\)). This holds for both Spearman and partial Spearman correlations (controlling for \(P\)).
\end{itemize}

\textbf{Inference:} The proposed bridging and breadth metrics are not merely proxies for productivity or impact. They capture orthogonal aspects of scholarly behavior related to knowledge integration and topical diversity, thereby validating their utility as distinct constructs in career trajectory analysis.

\subsubsection{Discriminant Validation on Labeled Groups}
\label{sec:discriminant_validation}

\textbf{Rationale:} A powerful test of a metric's utility is its ability to systematically distinguish between pre-defined, meaningful groups. Here, we test whether the proposed metrics can differentiate between comeback (CB) and dropout (DO) researchers, which is a central thesis of this work.

\textbf{Observation:} As detailed in Section~\ref{sec:results} and visualized in Figures~\ref{fig:communitybox}, \ref{fig:score_ci}, \ref{fig:gapentropy}, and \ref{fig:crosscomm}, the group contrasts are pronounced and statistically significant:
\begin{itemize}
    \item \textbf{Community Engagement \& Bridging:} Comeback authors cite a substantially broader range of communities (ACC: CB mean=74.78, DO mean=33.08) and hold higher bridging scores (\(B\): CB mean=0.608, DO mean=0.565). Bootstrap resampling confirms the mean differences are stable and their 95\% confidence intervals exclude zero (Fig.~\ref{fig:score_ci}).
    \item \textbf{Temporal Irregularity:} Comeback researchers exhibit significantly higher gap entropy (\(H_g\): CB mean=2.65, DO mean=1.52), confirming their more erratic publishing timelines (Fig.~\ref{fig:gapentropy}).
    \item \textbf{Statistical Significance:} Non-parametric Mann–Whitney U and Kolmogorov–Smirnov tests confirm these separations are highly significant (\(p < 10^{-24}\) for \(B\), \(p < 10^{-185}\) for ACC, \(p < 10^{-50}\) for \(H_g\)) with meaningful effect sizes (Table~\ref{tab:ttests_effects}).
\end{itemize}

\textbf{Inference:} The proposed metrics demonstrate high discriminant validity. They are not just different in theory; they empirically and robustly separate comeback researchers from dropouts, supporting the hypothesis that returnees have structurally and temporally distinct career patterns.

\subsubsection{Robustness Test}
\label{sec:robustness_test}

\textbf{Rationale:} The conclusions drawn from these metrics must not be fragile artifacts of specific analytical choices. Robustness testing ensures that the core findings hold under variations in methodology and parameters.

\textbf{Observation:} We subjected our analysis to multiple sensitivity checks, and the key conclusions remained stable:
\begin{itemize}
    \item \textbf{Community Detection:} The CB–DO contrasts in \(B\) and ACC persist under alternative community resolutions (Louvain \(\gamma \in [0.8, 1.4]\)) and when using the Leiden algorithm instead of Louvain.
    \item \textbf{Data Preprocessing:} The results are robust to the exclusion of self-citations, the use of different reference windows (\(\pm5\) years), and alternative binning strategies or logarithmic bases for calculating \(H_g\).
    \item \textbf{Predictive Criterion:} The out-of-sample predictive power of these metrics serves as a final, powerful robustness check. Classifiers using \(B\), ACC, and \(H_g\) consistently achieve high ROC–AUC (0.97, Fig.~\ref{fig:roc}), far outperforming a baseline model using only \(P\) and \(h\)-index (ROC–AUC \(\approx\) 0.54). SHAP analysis consistently ranks these proposed metrics as the most important features (Fig.~\ref{fig:shap}), confirming their stable and critical role in identifying comeback behavior.
\end{itemize}

\textbf{Inference:} The observed patterns are not methodological artifacts. The discriminative power and predictive utility of the proposed metrics are robust to a wide range of analytical decisions, strengthening the credibility of our findings and their applicability to real-world science policy and support systems.

% \textcolor{blue}{Comeback (CB) authors exceed dropouts (DO) on ACC and $B$/XCC, and exhibit higher $H_g$; non-parametric Mann–Whitney and Kolmogorov–Smirnov tests confirm significant separation with meaningful effect sizes (Figs.~\ref{fig:communitybox}, \ref{fig:crosscomm}, \ref{fig:gapentropy}). \textit{Robustness:} Conclusions persist under alternative community resolutions (Louvain $\gamma\in[0.8,1.4]$) and Leiden, exclusion of self-citations, $\pm 5$-year reference windows, and alternative binning/log bases for $H_g$. \textit{Criterion:} Out-of-sample classifiers using bridging/entropy features (ACC, $B$/XCC, $H_g$) achieve high ROC–AUC and substantially outperform a volume-only baseline; SHAP consistently ranks $H_g$, ACC, and $B$/XCC among the most informative predictors (Figs.~\ref{fig:roc}, \ref{fig:shap}). These results align the metric definitions in §\ref{sec:metrics} with the empirical patterns reported in §\ref{sec:results}.}

% \par \textcolor{red}{Please add the Metric Validation--- Which should include i) Statistical sanity check ii) Correlational validation with known metrics iii)Discriminant validation on labeled groups iv)robustness test}

\begin{figure}[ht]
\centering
\includegraphics[width=\linewidth, height=0.4\linewidth]{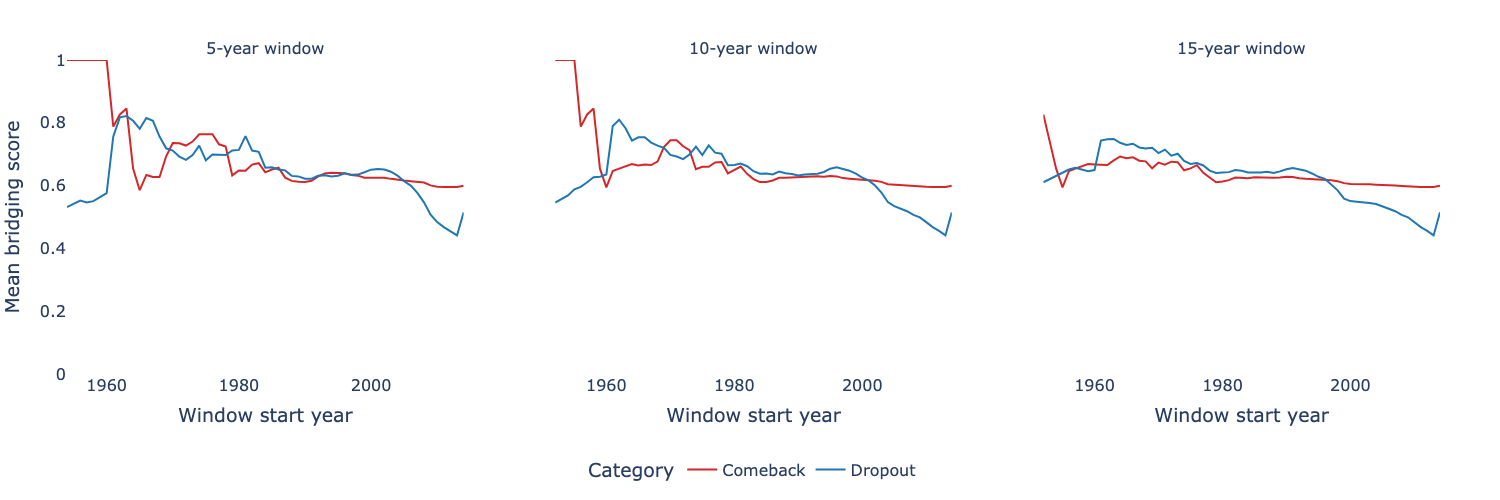}
\caption{Figure shows the moving window comparison of the average Bridging Score for Comeback vs. Dropout authors across 5, 10, and 15-year intervals. The x-axis likely represents the midpoint of the time window, and the y-axis represents the average Bridging Score.}
\label{fig:window}
\end{figure}

\begin{figure}[ht]
\centering
\includegraphics[width=0.8\linewidth]{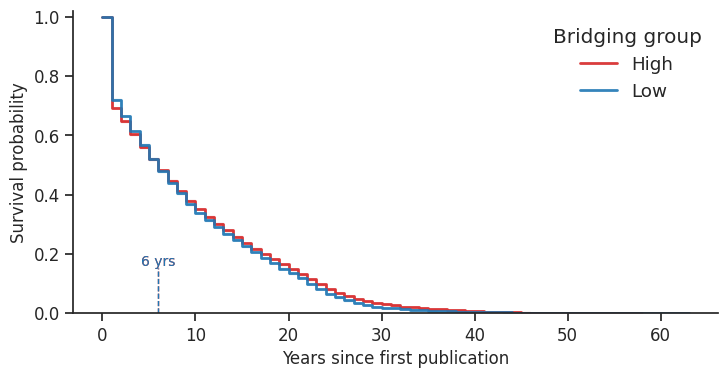}
\caption{The figure represents Kaplan-Meier survival curves showing the probability of academic persistence over time, stratified by a binary Bridging Score group (High vs. Low). The x-axis represents time (e.g., years), and the y-axis represents the survival probability.}
\label{fig:survival}
\end{figure}

These metrics jointly allow us to assess the structural and temporal distinctiveness of comeback researchers and support robust comparative and predictive modeling.

\section{Proposed Work}
\label{sec:methodology}

This section presents the comprehensive methodological framework developed to systematically investigate the structural roles and behavioral patterns of comeback researchers. Our approach is specifically designed to address the three core research questions outlined in Section~\ref{sec:intro}: (RQ1) cross-community knowledge transfer, (RQ2) structural bridging in citation networks, and (RQ3) predictive distinction between comeback and dropout researchers. 

To achieve this, we employ a multi-stage empirical pipeline that integrates large-scale bibliometric analysis with network science and machine learning. The methodology progresses from foundational data processing through sophisticated analytical modeling, ensuring each phase builds upon the previous to provide robust, interpretable insights. We begin with data collection and preprocessing from the AMiner dataset, followed by systematic author classification based on career discontinuity patterns. Subsequently, we construct dynamic citation networks and apply community detection to identify research fields, enabling the computation of both established and novel bibliometric indicators. Finally, we implement rigorous statistical testing and predictive modeling to validate our hypotheses and uncover the distinctive signatures of comeback researchers.

This integrated approach allows us to move beyond traditional bibliometric analysis and provide a network-theoretic understanding of how career interruptions can paradoxically enhance researchers' capacity for knowledge integration and interdisciplinary bridging.
\subsection{Dataset Summary}\label{s:dataset}

In this study, we utilize the AMiner dataset~\cite{aminerdataset}, a publicly available bibliometric corpus containing structured metadata on scientific publications, authorship, venues, and citation links. The dataset spans several decades and has been widely adopted in scholarly network analysis. For our experiments, we extracted and processed four main components: the paper dataset, author dataset, comeback author list, and dropout author list. Basic data schema and attributes are summarized in Table~\ref{table:dataset_format_comeback}.

\begin{table}[h!]
\centering
\begin{tabular}{||p{0.2\textwidth}|p{0.32\textwidth}||p{0.1\textwidth}|p{0.23\textwidth}||} 
 \hline
 \multicolumn{2}{||c||}{Paper} &
  \multicolumn{2}{c||}{Author}\\
 \hline
 Attributes & Attribute Definition & Attributes & Attribute Definition\\
 \hline\hline
 index & Paper ID & index & Author ID\\
 Paper Title & Title of the publication & Author & Author name \\
 Authors & List of authors (separated by `;') & Affiliation & Current affiliation \\
 Published Year & Year of publication & Papers & Total paper count \\
 Publication Venue & Conference or journal name & Citations & Total citation count \\
 Reference ID & List of cited paper IDs & H-index & Hirsch index \\
 Abstract & Abstract of the paper & pi, upi & Productivity indexes \\
 \hline
\end{tabular}
\caption{Data attributes extracted and processed from AMiner.}
\label{table:dataset_format_comeback}
\end{table}

The cleaned paper dataset contains over 2 million entries, with fields such as paper ID, title, authors, year, venue, reference list, and abstract. The author dataset includes more than 1.7 million author profiles with publication counts, citation counts, h-index values, affiliations, and inferred research interests. Additionally, curated lists of 1,425 comeback researchers and 11,351 dropout researchers were used to support comparative analysis.

\begin{table}[h!]
\centering
\begin{tabular}{ |c|c| } 
 \hline
 \textbf{Raw Dataset Summary} & \textbf{Value} \\
 \hline
 Total Papers Processed & 2,092,356 \\
 Total Authors Processed & 1,712,433 \\
 Valid Authors with Parsed Details & 1,132,637 \\
 Parsed Author-Paper Pairs & $\sim$4.7 million \\
 Reference Edges Extracted & $>$20 million \\
 \hline
\end{tabular}
\caption{Summary statistics of the raw parsed AMiner dataset.}
\label{tab:raw_dataset_stats}
\end{table}

\begin{table}[h!]
\centering
\begin{tabular}{ |c|c| } 
 \hline
 \textbf{Curated Researcher Statistics} & \textbf{Value} \\
 \hline
 Number of Comeback Authors & 1,425 \\
 Average Year Gap (Comeback) & 6.2 years \\
 Number of Dropout Authors & 11,351 \\
 Average Active Years (Dropouts) & 4.7 years \\
 Publication Span Considered & 1980---2014 \\
 \hline
\end{tabular}
\caption{Basic statistics of comeback and dropout researchers used in this study.}
\label{tab:curated_researchers}
\end{table}

The curated comeback set consists of authors who showed substantial early activity, had a discontinuity of at least 3 years, and later resumed publication. Each author record includes the starting year, discontinuation year, comeback year, and total gap duration. The dropout list consists of authors with no activity in the last 3+ years of the dataset, confirming their permanent exit. These author labels form the core of our comparative analysis pipeline and prediction experiments.

\subsection{Methodology}To systematically explore the knowledge transfer roles of comeback researchers, we adopt a multi-phase empirical pipeline built on large-scale bibliometric data. Our methodology integrates data preprocessing, career trajectory modeling, dynamic citation network analysis, community detection, semantic embedding-based topic modeling, and statistical classification using bridging and entropy-based metrics. This section elaborates each stage in detail, including mathematical formulations.

\begin{figure}[ht]
    \centering
    \includegraphics[width=\linewidth, height=0.4\linewidth]{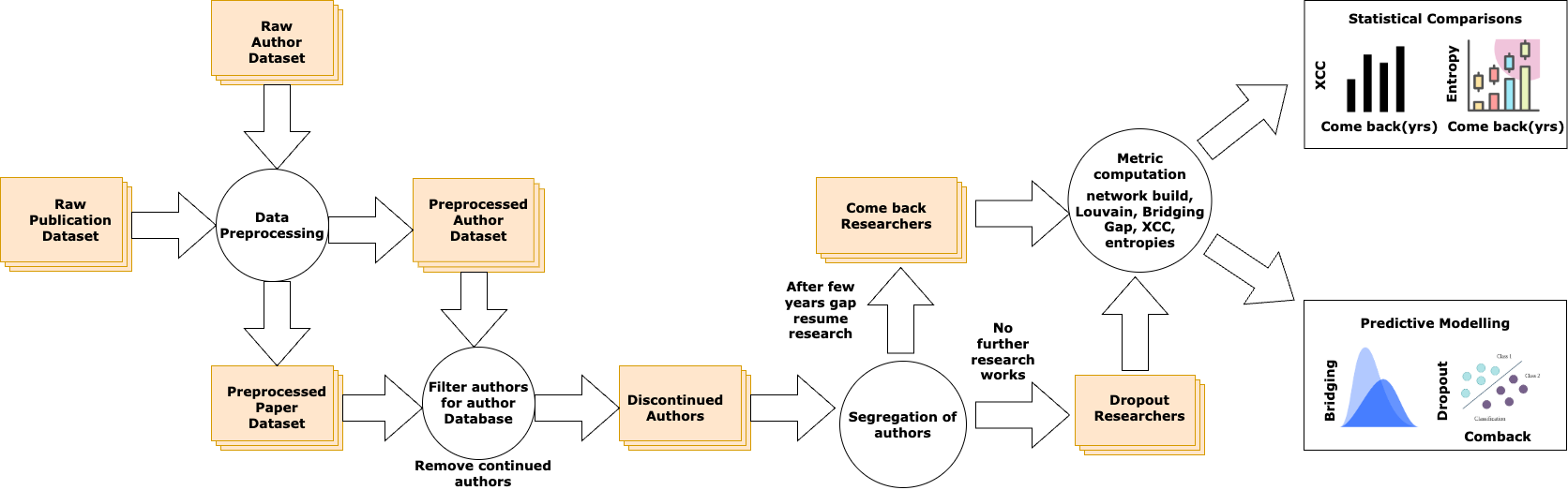}
    \caption{The figure represents the schematic overview of the methodological workflow. The diagram shows the pipeline from data preprocessing and author segmentation to network analysis, metric computation, and final statistical and predictive evaluation.}
    \label{fig:methodology}
\end{figure}

\par Figure~\ref{fig:methodology} illustrates the entire workflow. Raw author and publication datasets are preprocessed to isolate discontinued researchers. These are then segregated into comeback and dropout cases based on return-to-publication status. We compute metrics like bridging centrality and entropy, followed by comparative and predictive analyses to understand structural roles.

\subsubsection{Data Collection and Preprocessing}Our analysis is grounded in the Aminer academic graph dataset, which includes metadata for millions of academic publications, including paper titles, authors, years, and reference lists. We begin by parsing the dataset to retain only entries with complete information. The author lists are tokenized and exploded into individual rows to construct tuples of the form $(a, p, y)$, where $a$ is an author, $p$ a paper ID, and $y$ the year of publication.

We then standardize all reference lists by replacing delimiters (e.g., commas and semicolons) with a consistent separator, and parse citation edges such that for a given paper $p_i$ citing a set of references $R(p_i)$, we generate directed edges $(p_i \rightarrow p_j)$ for all $p_j \in R(p_i)$. This results in a yearly citation graph $G_t = (V_t, E_t)$, where $V_t$ represents papers up to year $t$, and $E_t$ is the set of directed citation edges.

\subsubsection{Author Career Classification}

To classify researchers, we use their publication timelines derived from the preprocessed data. Let $P_a = \{y_1, y_2, \ldots, y_n\}$ represent the sorted publication years of author $a$. We define the largest inter-publication gap as:

\[
G(a) = \max_i (y_{i+1} - y_i)
\]

An author is labeled a \textbf{comeback} (CB) if $G(a) \geq 3$ and there exists at least one publication after the gap. A \textbf{dropout} (DO) is defined as someone with $G(a) \geq 3$ and no subsequent publications for the remaining time window (at least 3 years before the dataset's end). Authors with no significant publishing gap ($G(a) < 3$) are categorized as \textbf{active} (AC). This temporal filtering is used to study re-entry behavior and publication inactivity.

\subsubsection{Statistical Modeling}
\label{sec:stat_modeling}To rigorously validate the distinctiveness of comeback researchers, we employ a statistical modeling framework designed to test for significant differences between the comeback (CB) and dropout (DO) cohorts. The primary objective is to ascertain whether the observed disparities in network-based and temporal metrics are statistically meaningful and not attributable to random chance. This forms the foundational empirical evidence for addressing our core research questions (RQ1 and RQ2). Our statistical analysis proceeds as follows, with each author treated as an independent unit of observation:
\begin{itemize}
    \item \textbf{Hypothesis Testing for Group Differences:} We conduct non-parametric tests to compare the distributions of all proposed and baseline metrics (e.g., Bridging Score \(B\), ACC, \(H_g\), publication count \(P\)) between the CB and DO groups. The Mann–Whitney U test is used to assess whether one group tends to have larger values than the other, while the Kolmogorov–Smirnov (KS) test determines if the underlying distributions of the two groups differ significantly. To quantify the magnitude of these differences, beyond mere statistical significance, we report effect sizes: Cliff's \(\delta\) for the Mann–Whitney test and the KS statistic \(D\) for the Kolmogorov–Smirnov test. Given the multiple comparisons across different metrics, we control the False Discovery Rate (FDR) by applying the Benjamini–Hochberg procedure to the obtained \(p\)-values.

    \item \textbf{Correlational Analysis:} To understand the relationships between our proposed metrics and to check for convergent and divergent validity, we estimate Spearman's rank correlation coefficients. Specifically, we examine the correlation between the edge-weighted Bridging Score (\(B\)) and the paper-weighted Cross-Community Citation Rate (XCC) to verify they capture a similar construct (convergent validity). We further compute partial Spearman correlations, controlling for total publications (\(P\)), to ensure that the bridging and diversity metrics provide information orthogonal to simple productivity measures (divergent validity).

    \item \textbf{Ensuring a Fair Comparison:} A critical aspect of our methodology is the use of a \textit{non-leaky observation window}. For comeback authors, features are computed using only their publications from their first paper up to the year immediately preceding their return. For dropout authors, we use a matched-length career window ending at their last active year. This ensures we are comparing early-career signals and not the consequences of the comeback event itself. To further reinforce the reliability of our key findings, we derive bootstrap confidence intervals for the mean differences in critical metrics like Bridging Score and Community Count between the CB and DO cohorts.
\end{itemize}

This comprehensive statistical approach allows us to move beyond anecdotal observation and provide quantitative, robust evidence for the unique structural and temporal signatures of comeback researchers.

% \subsubsection{Statistical Modeling}
% \textcolor{blue}{To assess group-level differences, we treat each author as a unit of analysis and compare comeback versus dropout cohorts on all network and diversity metrics using non-parametric tests (Mann--Whitney U and Kolmogorov--Smirnov). For interpretability, we report effect sizes (Cliff's $\delta$ for Mann--Whitney; KS statistic $D$ for Kolmogorov--Smirnov) and adjust $p$-values for multiple comparisons via Benjamini--Hochberg control of the false discovery rate. Where relevant, we estimate Spearman and partial Spearman correlations (controlling for total publications $P$) between bridging score ($B$) and cross-community citation rate (XCC) to verify convergent and divergent relationships with established measures. All tests use features computed from a non-leaky observation window (pre-return for comebacks and a matched career window for dropouts), and we derive bootstrap confidence intervals for key contrasts.}

\subsubsection{Predictive Modeling}
\label{sec:predictive_modeling}Beyond establishing statistical differences, a critical test of the practical utility of our proposed metrics is their ability to prospectively \textit{identify} researchers with the potential to return after a hiatus. Therefore, we develop a predictive modeling framework to answer Research Question 3 (RQ3): Can we reliably distinguish future comeback authors from permanent dropouts? Success in this task would not only validate the discriminative power of our features but also pave the way for data-driven tools to support early intervention and targeted institutional policies for researchers on non-linear career paths. Our predictive modeling pipeline is designed for robustness, interpretability, and real-world applicability:
\begin{itemize}
    \item \textbf{Model Selection and Training:} We train a diverse set of powerful, non-linear classifiers, including Random Forest, XGBoost, and LightGBM, known for their strong performance on tabular data. To ensure a rigorous evaluation, we employ \textit{stratified 5-fold cross-validation} with an author-level split, guaranteeing that all papers from a single author remain within either the training or test set in each fold, thus preventing data leakage. Given the inherent class imbalance between comeback and dropout researchers, we utilize class-weighted loss functions to ensure the models learn from both classes effectively.

    \item \textit{Feature Sets for Ablation Study:} To isolate the contribution of our novel metrics, we conduct a comparative ablation study using two distinct feature sets:
    \begin{enumerate}
        \item \textbf{Baseline Feature Set:} This set comprises traditional bibliometric indicators—total paper count (\(P\)) and the h-index (\(h\))—serving as a benchmark to represent conventional evaluation criteria.
        \item \textbf{Bridging \& Entropy Feature Set:} This is our proposed set, containing the novel metrics introduced in Section~\ref{sec:proposed_metrics}: Cross-Community Citation Rate (XCC), Bridging Score (\(B\)), Authored-Community Count (ACC), and Gap Entropy (\(H_g\)).
    \end{enumerate}
    The performance gap between models trained on these two sets directly quantifies the added value of our proposed features.

    \item \textbf{Performance Evaluation and Interpretation:} We evaluate model performance using a comprehensive suite of metrics: Accuracy, Precision, Recall, F1-score, and the Area Under the Receiver Operating Characteristic Curve (ROC-AUC). Probability calibration via Platt scaling is applied for reliable threshold-based analysis. To demystify the ``black box'' nature of ensemble models and provide \textit{criterion validity} for our proposed metrics, we compute SHAP (SHapley Additive exPlanations) values. This allows us to globally rank feature importance and understand the marginal contribution of each feature (e.g., \(H_g\), ACC, \(B\)) to the model's prediction, thereby linking model decisions back to our theoretical framework.
\end{itemize}

This predictive framework transforms our theoretical insights into a practical tool, testing whether the structural and temporal signatures of comeback researchers are not only significant but also sufficiently patterned to be predictable.

% \subsubsection{Predictive Modeling}
% \textcolor{blue}{Separately, we build out-of-sample classifiers to predict comeback versus dropout status from author features. We train Random Forest, XGBoost, and LightGBM under stratified 5-fold CV with an author-level split and class-weighted losses. Features include ACC, $B$, XCC, and $H_g$ computed in the non-leaky observation window. Performance is reported via Accuracy, Precision, Recall, F1, and ROC–AUC. Two feature sets are compared in ablation: a \emph{baseline} with total paper count ($P$) and $h$-index, and a \emph{bridging} set comprising XCC, bridging score $B$, community count ACC, and gap entropy $H_g$. Performance is evaluated using accuracy, precision, recall, F1, and ROC~AUC, with probability calibration (Platt scaling) for threshold analyses. To explain model decisions and provide criterion validity for the proposed metrics, we compute SHAP values (global rankings and fold-stable importances).}

\section{Results and Analysis}
\label{sec:results}

This section presents a comprehensive analysis of the structural, temporal, and predictive characteristics of comeback researchers, systematically addressing our three research questions. Through rigorous statistical testing, network analysis, and machine learning evaluation, we demonstrate the distinct bridging behavior and predictable patterns of researchers who return to academia after prolonged hiatus.

\begin{table}[ht]
\centering
\renewcommand{\arraystretch}{1.15}
\setlength{\tabcolsep}{3pt}
\small
\begin{tabularx}{\linewidth}{|l|c|c|>{\centering\arraybackslash}X|>{\centering\arraybackslash}X|>{\centering\arraybackslash}X|}
\hline
\textbf{Metric} & \textbf{CB mean} & \textbf{DO mean} & \textbf{Mean diff [95\% CI]} & \textbf{$t$ (CB–DO), $p$} & \textbf{Effects (d/$\delta$)} \\
\hline
Bridging $B$ & 0.608 & 0.565 & 0.044 [0.035, 0.052] & 10.08,\ $7.84\!\times\!10^{-24}$ & 0.131 / 0.016 \\
ACC (communities) & 74.78 & 33.08 & 41.70 [38.88, 44.57] & 29.34,\ $4.33\!\times\!10^{-185}$ & 0.402 / 0.616 \\
\hline
\end{tabularx}
\caption{Statistical comparison of the Bridging Score (B) and Authored-Community Count (ACC) between Comeback (CB) and Dropout (DO) cohorts within the observation window. The table reports group means, mean differences with 95\% confidence intervals, Welch's t-test results (t-statistic and p-value), and effect sizes (Cohen's $d$ / Cliff's $\delta$)}
\label{tab:ttests_effects}
\end{table}

\subsection{Bridging Behavior and Structural Diversity}
\label{sec:bridging_behavior}To address RQ1 (cross-community knowledge transfer) and RQ2 (structural bridging in citation networks), we first examine whether comeback researchers engage with a broader range of research communities and occupy more connective positions compared to dropout and active researchers. We evaluate two key metrics: \textit{Community Count (ACC)}, representing the number of distinct communities cited or engaged with, and \textit{Bridging Score (B)}, reflecting the fraction of an author's outgoing citations that cross community boundaries.

\par Our analysis reveals that comeback researchers cite a substantially broader range of communities (mean: 74.78) compared to dropout researchers (33.08) and active researchers (10.70). Similarly, the average bridging score for comeback authors is higher (0.608) than both dropouts (0.565) and actives (0.579). Figure~\ref{fig:communitybox} visualizes these distributions, showing comeback authors with both higher medians and wider spreads in community engagement and bridging roles. Statistical tests confirm these differences are highly significant (Table~\ref{tab:ttests_effects}), with bootstrap confidence intervals for mean differences excluding zero (Figure~\ref{fig:score_ci}). The substantial effect sizes (Cliff's $\delta$ = 0.616 for ACC, 0.016 for B) indicate meaningful practical differences beyond statistical significance\footnote{\small \textit{Note:} The last column in Table~\ref{tab:ttests_effects} reports effect sizes as (Cohen's $d$ / Cliff's $\delta$). For $B$, the difference is statistically significant but \emph{practically modest} ($d\!\approx\!0.13$, $\delta\!\approx\!0.016$), whereas ACC shows a \emph{large} practical difference ($d\!\approx\!0.40$, $\delta\!\approx\!0.616$).}.

\par These results provide strong affirmative answers to both RQ1 and RQ2. Comeback researchers not only engage with more diverse research communities but also consistently occupy structural positions that facilitate knowledge flow across disciplinary boundaries, functioning as true bridges rather than merely broad participants.

\subsection{Cross-Community Citation Patterns}
\label{sec:citation_patterns}To further investigate RQ1 regarding cross-community knowledge transfer, we examine the \textit{Cross-Community Citation Rate (XCC)}, which measures the proportion of citations directed outside an author's own community. We compare XCC distributions between comeback and dropout cohorts using statistical testing and visualization.

\par As shown in Figure~\ref{fig:crosscomm}, comeback researchers maintain significantly higher median cross-community citation rates. A two-sample t-test confirms this observation with high significance ($t = 10.08$, $p < 10^{-23}$). The near-zero Spearman correlation ($\rho \approx 0.004$) between ACC (breadth) and B (edge-weighted cross-share) indicates these metrics capture complementary facets of cross-community engagement. Qualitative analysis of topical breadth through sunburst visualization (Figure~\ref{fig:topic_sunburst}) further supports these findings, showing comeback authors cover a wider spread of high-level topics with a heavier tail of specific terms.

\par These results reinforce our answer to RQ1, demonstrating that comeback researchers' cross-community engagement extends beyond mere breadth to include active knowledge importation from external domains, consistent with their role as interdisciplinary integrators.

\begin{figure}[ht]
\centering
\includegraphics[width=\linewidth]{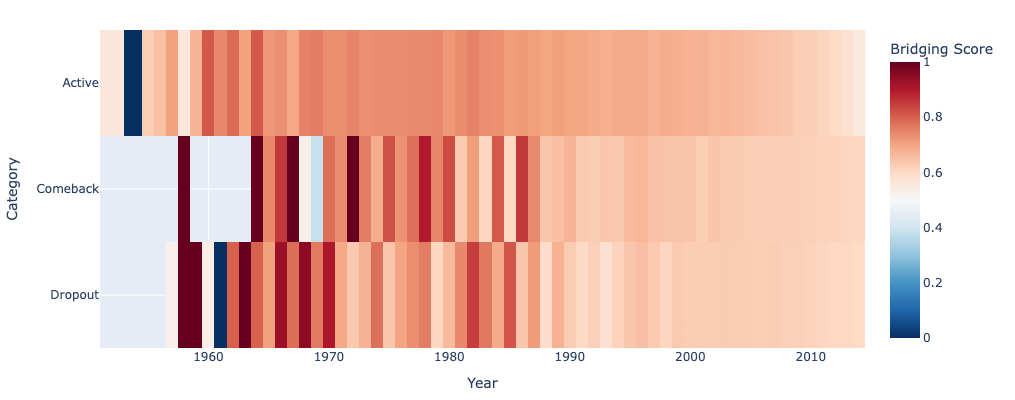}
\caption{The figure represents heatmap of the average Bridging Score across publication years, segmented by author category (Comeback, Dropout, Active). The x-axis represents the year, and the y-axis represents the author category. The color intensity corresponds to the average Bridging Score value.}
\label{fig:heatmap}
\end{figure}

\subsection{Temporal Dynamics of Bridging}
\label{sec:temporal_dynamics}To understand how bridging behavior evolves over time and whether it correlates with re-entry patterns, we examine the temporal dynamics of bridging activity. We construct heatmaps of yearly average bridging scores and perform sliding window analyses across 5, 10, and 15-year spans.

\par The temporal heatmap (Figure~\ref{fig:heatmap}) reveals that comeback authors exhibit bursts of high bridging activity centered around their return years. Active researchers maintain consistent but moderate bridging, while dropouts show gradual decline. Sliding window analysis (Figure~\ref{fig:window}) shows comeback researchers outperform dropouts in 55\% of 5-year windows, with this advantage diminishing in larger windows.

\par These temporal patterns suggest that bridging is not merely a stable trait but a strategic behavior concentrated around re-entry periods. The episodic nature of high bridging activity supports the hypothesis that comeback researchers actively leverage network connectivity as a reintegration mechanism.

\subsection{Publication Irregularity and Career Patterns}
\label{sec:career_irregularity}To characterize the temporal signatures of non-linear careers, we examine publication regularity through gap entropy analysis. We compute Gap Entropy ($H_g$) for each author, quantifying the irregularity of publication timelines.

\par Comeback researchers exhibit significantly higher gap entropy (mean: 2.65) than dropouts (mean: 1.52), with $t = 48.46$ and $p < 10^{-50}$ (Figure~\ref{fig:gapentropy}). Survival analysis further reveals that high-bridging authors demonstrate marginally longer academic persistence ($\chi^2 = 477.93$, $p < 10^{-100}$).

\par The elevated gap entropy confirms that career volatility is a hallmark of comeback behavior. Rather than indicating academic weakness, this irregularity appears to coincide with structurally significant re-engagements, challenging conventional metrics that prioritize steady output.

\begin{figure}[t]
  \centering
  \includegraphics[width=\linewidth, height=0.6\linewidth]{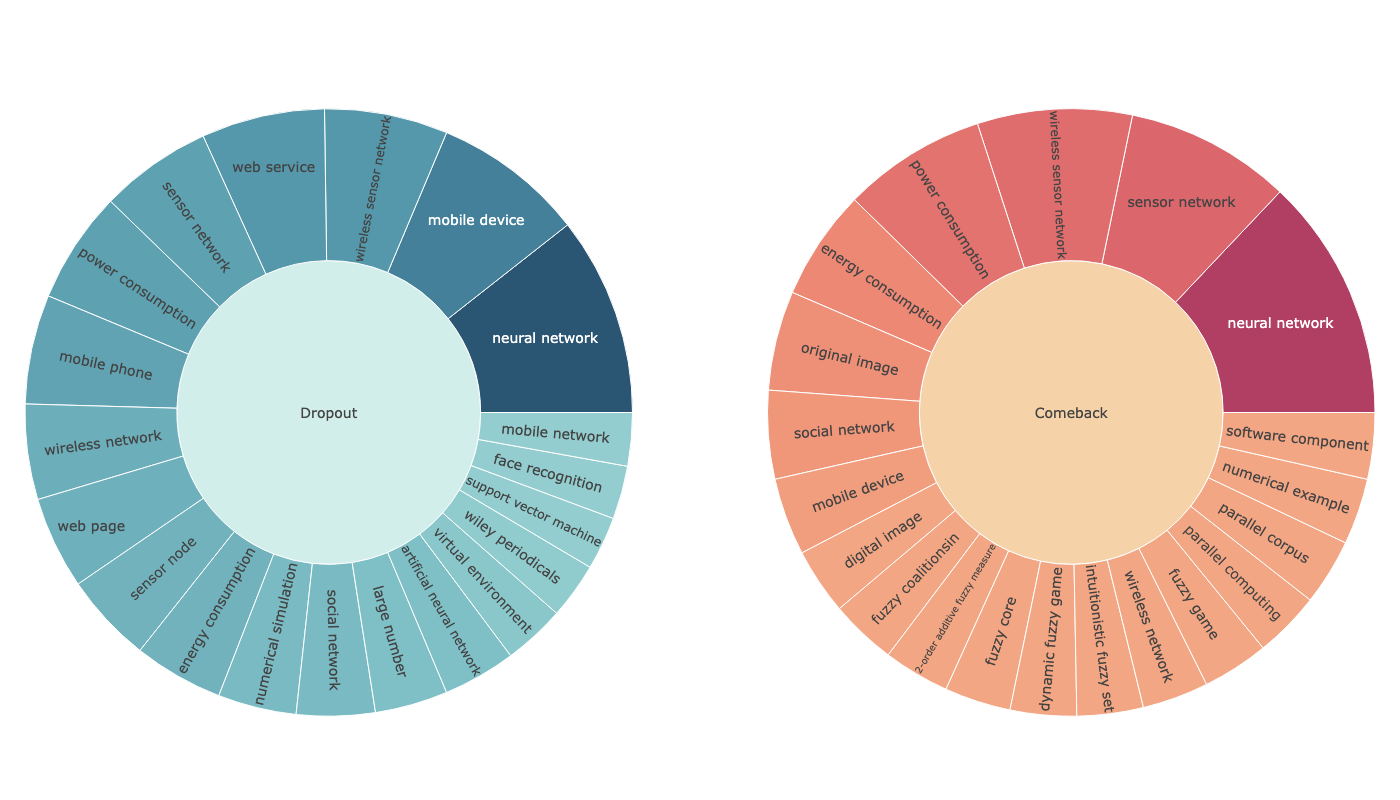}% or your final filename
  \caption{Figure represents the sunburst visualization of the topic mix for each author cohort. Each ring aggregates the top subject terms by frequency within the observation window, with segment size proportional to term frequency.}
  \label{fig:topic_sunburst}
\end{figure}

\begin{table}[ht]
\centering
\setlength{\tabcolsep}{5.85pt}
\small
\begin{tabularx}{\linewidth}{|l|c|c|c|c|c|}
\hline
\textbf{Model} & \textbf{Accuracy} & \textbf{Precision} & \textbf{Recall} & \textbf{F1} & \textbf{ROC--AUC} \\
\hline
Random Forest       & 0.927 & 0.974 & 0.888 & 0.929 & 0.972 \\
XGBoost             & 0.911 & 0.971 & 0.858 & 0.911 & 0.965 \\
LightGBM            & 0.906 & 0.959 & 0.862 & 0.908 & 0.963 \\
KNN (k=5)           & 0.900 & 0.950 & 0.859 & 0.902 & 0.943 \\
Gradient Boosting   & 0.871 & 0.943 & 0.809 & 0.871 & 0.927 \\
SVM (RBF)           & 0.735 & 0.802 & 0.672 & 0.731 & 0.792 \\
Logistic Regression & 0.731 & 0.739 & 0.768 & 0.753 & 0.756 \\
\hline
\end{tabularx}
\caption{The Table shows the out-of-sample performance metrics for various classification models predicting comeback status, using the bridging and entropy feature set (ACC, B, XCC, $H_g$). The metrics reported are Accuracy, Precision, Recall, F1-score, and ROC-AUC, obtained via stratified cross-validation.}
\label{tab:model_perf}
\end{table}

\subsection{Predictive Distinction of Comeback Researchers}
\label{sec:predictive_modeling}To address RQ3 regarding the reliable distinction between comeback and dropout researchers, we develop predictive models using our proposed bridging and entropy metrics. We train multiple classifiers using two feature sets: (1) traditional bibliometric indicators (baseline), and (2) our proposed bridging and entropy metrics.

\par
\begin{figure}[ht]
\centering
\includegraphics[width=0.8\textwidth]{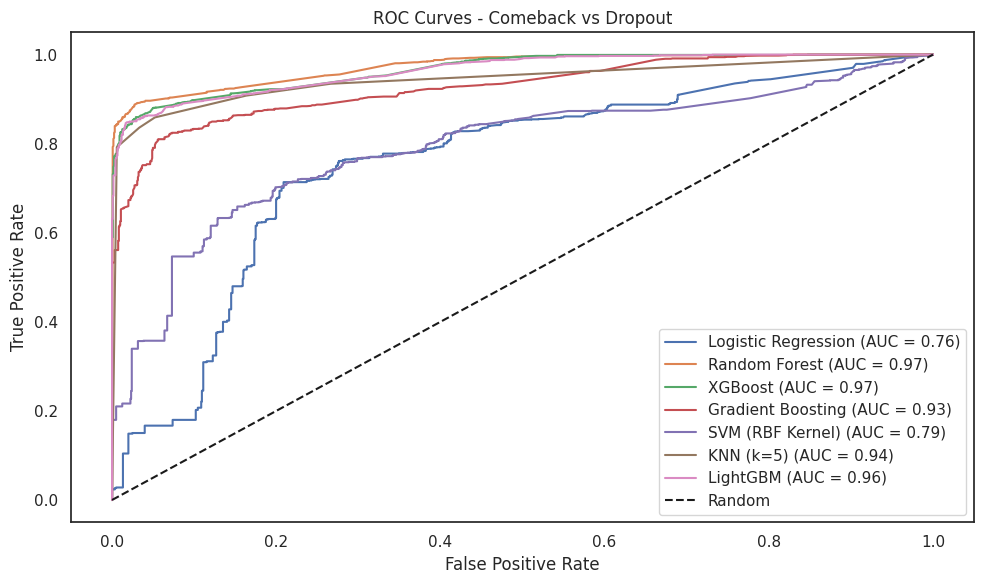}
\caption{Figure shows the receiver Operating Characteristic (ROC) curves for various machine learning models predicting comeback status. The x-axis represents the False Positive Rate, and the y-axis represents the True Positive Rate. The diagonal dashed line represents a random classifier.}
\label{fig:roc}
\end{figure}

\begin{figure}[ht]
\centering
\includegraphics[width=0.8\linewidth]{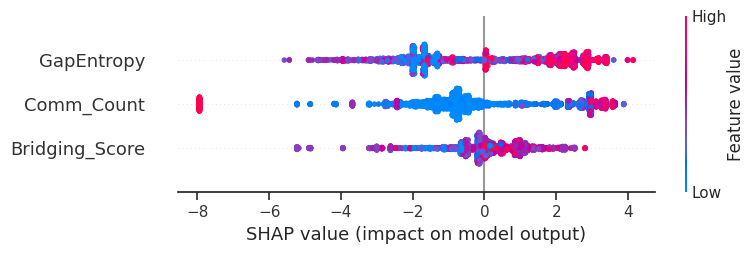}
\caption{The figure represents the SHAP summary plot showing the impact of the top features on the model's output. Each point represents a researcher instance. The x-axis is the SHAP value (impact on model output), and the y-axis lists the features. The color represents the feature value from low (blue) to high (red).}
\label{fig:shap}
\end{figure}

 Ensemble methods achieve exceptional performance, with Random Forest and XGBoost reaching ROC-AUC of 0.97, indicating excellent separability—well above the logistic baseline (0.76) and the volume-only baseline (≈ 0.54).(Figure~\ref{fig:roc}, Table~\ref{tab:model_perf}). In stark contrast, a baseline model using only publication count and h-index yields near-chance performance (ROC-AUC = 0.54). SHAP analysis (Figure~\ref{fig:shap}) consistently ranks Gap Entropy ($H_g$) as the dominant driver, followed by Community Count (ACC) and Bridging Score (B). It identifies that higher values of these features push predictions toward ‘comeback’, while lower values push toward ‘dropout’.

\par These results provide a strong affirmative answer to RQ3. The superior predictive performance demonstrates that bridging-based features capture distinctive signatures of comeback trajectories that are invisible to traditional metrics. The high feature importance of our proposed metrics validates their utility for early identification and targeted support. Qualitative analysis of topical breadth through sunburst visualization (Figure~\ref{fig:topic_sunburst}) further supports these findings, showing comeback authors cover a wider spread of high-level topics with a heavier tail of specific terms.

% Notably, a baseline model using only total publication count and h-index yields an ROC-AUC of 0.54. This highlights the inadequacy of standard metrics in identifying comeback trajectories and reinforces the predictive utility of our bridging-based features (supporting RQ3).

\begin{table}[ht]
\centering
\renewcommand{\arraystretch}{1.15}
\setlength{\tabcolsep}{4pt}
{\small
\begin{tabularx}{\textwidth}{|p{3.6cm}|>{\raggedright\arraybackslash}X|p{3.1cm}|p{2.2cm}|}
\hline
\textbf{Metric / Model} & \textbf{CB vs.\ DO (or Score)} & \textbf{Primary test / stat} & \textbf{Where shown} \\
\hline
\multicolumn{4}{|l|}{\textbf{A. CB–DO Metric contrasts}} \\
\hline
Distinct authored communities (ACC) & CB: 74.78 (mean); DO: 33.08 (mean); CI excludes 0 & Bootstrap mean diff (95\% CI $>0$) & Figs.~\ref{fig:communitybox}, \ref{fig:score_ci} \\
\hline
Bridging score ($B$) & CB: 0.608 (mean); DO: 0.565 (mean); CI excludes 0 & Bootstrap mean diff (95\% CI $>0$) & Figs.~\ref{fig:communitybox}, \ref{fig:score_ci} \\
\hline
Cross-community citation (XCC) & CB higher than DO & $t=10.08$, $p<10^{-23}$ & Fig.~\ref{fig:crosscomm} \\
\hline
Gap entropy ($H_g$) & CB: 2.65 (mean); DO: 1.52 (mean) & $t=48.46$, $p<10^{-50}$ & Fig.~\ref{fig:gapentropy} \\
\hline
Kaplan-Meier survival (shows persistence)  & CB shows higher survival & log-rank $\chi^2=477.93$, $p<10^{-100}$ & Fig.~\ref{fig:survival} \\
\hline
\multicolumn{4}{|l|}{\textbf{B. Predictive (criterion) validation}} \\
\hline
RF / XGB (bridging+entropy features) & ROC–AUC $=0.97$; SHAP: $H_g$, ACC, B/XCC top & Calibrated; stratified CV & Figs.~\ref{fig:roc}, \ref{fig:shap} \\
\hline
Logistic regression (same features) & ROC–AUC $=0.76$ & Interpretable baseline & Fig.~\ref{fig:roc} \\
\hline
Baseline ($P,h$ only) & ROC–AUC $=0.54$ & Near-chance (criterion check) & Fig.~\ref{fig:roc} \\
\hline
\end{tabularx}
}
\caption{Metrics \& validation scores summarizing Comeback (CB)–Dropout (DO) contrasts and predictive performance, with statistical tests and figure references.}
\label{tab:metrics_scores}
\end{table}

\subsection{Integrated Discussion and Implications}
\label{subsec:discussion}
This section synthesizes the empirical findings from our analyses into a cohesive interpretation, drawing out their broader significance for the science of science, academic practice, and research policy. The rationale for this integrated discussion is to move beyond reporting statistical results and to articulate a coherent narrative about the unique role of comeback researchers. We have established that they are structurally and behaviorally distinct; here, we explore the \textit{why} and \textit{so what} of these differences. Specifically, we will first consolidate the evidence to describe the general phenomenon of the comeback researcher. We will then discuss the theoretical implications of our findings, challenging established models of scientific careers. Finally, we will translate these insights into actionable practical strategies and concrete policy recommendations, arguing for a systemic re-evaluation of how non-linear academic paths are perceived and supported.
\subsubsection{General Discussion}

The convergent evidence across all analyses reveals that comeback researchers are structurally distinct, temporally adaptive, and highly predictable actors in scientific networks. A central finding is their broad community engagement; as shown in Figure~\ref{fig:communitybox}, they cite a significantly larger number of distinct communities and maintain higher bridging scores than both dropouts and active peers. This suggests their return is characterized by a strategic re-engagement with a diverse scholarly landscape, positioning them as integrative agents rather than passive rejoiners.

This role is further supported by their citation behavior. Our analysis demonstrates that comeback researchers direct a higher proportion of their citations toward external communities (Figure~\ref{fig:crosscomm}), actively operating at disciplinary boundaries where knowledge transfer is most valuable. Qualitative analysis of topical breadth through sunburst visualization (Figure~\ref{fig:topic_sunburst}) reinforces this, showing comeback authors cover a wider spread of high-level topics.

These structural advantages manifest in distinct temporal patterns. The bridging activity of comeback researchers peaks around their return year (Figure~\ref{fig:heatmap}), indicating a burst-like, strategic effort to reintegrate. This episodic bridging, coupled with significantly higher gap entropy (Figure~\ref{fig:gapentropy}), confirms that their careers are non-linear, marked by pauses and surges that correlate with structurally significant re-engagements.

Critically, these behavioral signatures are highly predictable. Classifiers using our proposed bridging and entropy features achieve exceptional performance (ROC-AUC of 0.97, Figure~\ref{fig:roc}), far outperforming models based on traditional metrics like publication count and h-index (ROC-AUC $\approx$ 0.54). SHAP analysis (Figure~\ref{fig:shap}) confirms that Gap Entropy ($H_g$), Community Count (ACC), and Bridging Score ($B$) are the most impactful predictors. A comprehensive summary of these results is provided in Table~\ref{tab:metrics_scores}, underscoring the robustness of our findings.

\subsubsection{Theoretical Implications}

Our findings challenge foundational theories in the science of science. The \textbf{cumulative advantage (Matthew Effect)} model \cite{Merton1988TheME}, which posits that continuous productivity begets further success, is insufficient to explain the comeback phenomenon. We demonstrate that strategic discontinuity, followed by re-entry, can paradoxically enhance a researcher's capacity for boundary-spanning and knowledge integration.

The observed ``\textbf{bridge-on-return}'' pattern suggests that career interruptions can serve as a crucible for refining scholarly focus and fostering eclectic, interdisciplinary engagements. This positions comeback researchers as key agents in filling ``structural holes'' within citation networks, facilitating novel knowledge recombinations in a way that challenges models prioritizing uninterrupted, linear career progression.

\subsubsection{Practical Implications}

The predictive utility of our metrics enables a shift from retrospective assessment to proactive support. The high feature importance of Gap Entropy, ACC, and Bridging Score provides a data-driven foundation for:

\begin{itemize}
    \item \textbf{Early Identification:} Institutions and mentors can use these interpretable features to identify early-career researchers with high comeback potential after a hiatus, enabling timely intervention.
    \item \textbf{Targeted Support:} This identification can facilitate targeted support mechanisms such as re-entry fellowships, bridge funding, and dedicated mentorship programs designed for researchers on non-linear paths.
    \item \textbf{Fairer Evaluation:} For hiring and promotion committees, these metrics offer tools to recognize and value the unique integrative contributions of researchers with career gaps, moving beyond a narrow focus on continuous productivity.
\end{itemize}

\subsubsection{Policy Recommendations}

To foster a more inclusive and resilient scientific ecosystem, we advocate for policy changes that recognize the value of non-linear careers:

\begin{enumerate}
    \item \textbf{Integrate Network-Based Indicators:} Funding agencies and academic institutions should complement traditional bibliometric indicators (e.g., h-index, publication count) with network-based metrics of knowledge integration and bridging in their evaluation criteria.
    \item \textbf{Create Re-entry Funding Lines:} Dedicated grant programs and fellowships should be established specifically for researchers returning after a prolonged career break, reducing the financial and professional barriers to re-entry.
    \item \textbf{Promote Mentorship and Community:} Develop institutional policies that create formal mentorship pipelines and community-building initiatives for comeback researchers, helping them navigate the challenges of reintegration and leverage their unique cross-disciplinary potential.
\end{enumerate}

The inter-dependencies between bridging behavior, diversity of engagement, and predictability illustrate that the comeback phenomenon is a multifaceted engine for scientific connectivity. Supporting these researchers through informed policies and practices can strengthen the structural integrity of the entire knowledge ecosystem.

\section{Discussion and Conclusions}
\label{sec:conclusion}This study systematically characterized ``comeback researchers''—those returning to academia after a hiatus of three or more years—through a large-scale bibliometric analysis of the AMiner dataset, encompassing 113,637 early-career researchers, of which 1,425 were curated as comeback cases. Our findings demonstrate that these individuals do not merely resume previous trajectories but engage in strategically distinct behaviors, citing across a substantially broader range of research communities (126\% more than dropouts) and exhibiting a 7.6\% higher Bridging Score, which underscores their pronounced role as knowledge bridges between otherwise disconnected scientific silos—a role further evidenced by their preference for fast-cycle, boundary-crossing publication venues. Temporally, their careers are marked by significantly higher Gap Entropy (74\% higher than dropouts), reflecting strategic, non-linear publication patterns where bridging activity peaks specifically around their return years, suggesting a deliberate reintegration mechanism. Crucially, these structural and temporal signatures are highly predictable, with models using our novel bridging- and entropy-based metrics achieving an ROC-AUC of 97\%, far outperforming models based on traditional metrics like publication count and h-index (ROC-AUC = 54\%), thereby confirming that the value of comeback researchers is invisible to conventional, volume-based evaluation frameworks and providing the first large-scale, network-theoretic evidence that career discontinuity can paradoxically reposition researchers as pivotal agents of interdisciplinary knowledge transfer and structural cohesion within science.

\par While this study offers novel insights, it is subject to several limitations inherent to its data and methodological choices. The analysis is grounded in the AMiner bibliographic dataset, which, while extensive, may have coverage biases affecting generalizability. The operational definition of a ``comeback researcher'' is based exclusively on observable publication gaps and does not capture the underlying reasons for the hiatus (e.g., personal leave, industry work) or the qualitative experiences during the break, limiting a richer understanding of reintegration mechanisms. The community detection via the Louvain algorithm provides a data-driven but abstract notion of ``research community'', whose stability and interpretability are sensitive to resolution parameters. Further, the predictive models, while highly accurate, are trained on historical data and require continuous validation as scientific norms evolve. 
\par Consequently, this work opens several promising avenues for future research, including validating the generalizability of our findings across different disciplines, countries, and institutional contexts. Particularly, the observed geographic patterns (e.g., higher comeback shares in China and the U.S.) suggests that structural and cultural factors significantly influence re-entry patterns and warrant deeper investigation. Future work should move beyond publication metadata to incorporate qualitative and survey-based methods to understand the subjective experiences, challenges, and strategic decisions of comeback researchers, thereby explaining the \textit{why} behind the observed structural patterns. It can be extended the predictive framework into real-time, ethically deployed decision-support systems for funders and institutions to identify and provide targeted support, such as re-entry fellowships and mentorship. The exploration of  the semantic content of the research produced before and after their hiatus using text mining and NLP techniques could reveal the precise nature of the knowledge transferred and the thematic shifts undergone. By pursuing these directions, we can build more inclusive and accurate models of scientific careers that recognize the latent value in non-linear paths and strengthen the entire knowledge ecosystem.

\section*{Declarations}
\begin{itemize}
\item \textbf{Funding}:
No funds, grants, or other support were received to conduct this study.
\item \textbf{Conflict of interest/Competing interests}:  On behalf of all authors, the corresponding author states that there is no conflict of interest.
\item \textbf{Data availability}: Publicly available.
\item \textbf{Author contribution}: {\bf Somyajit Chakraborty:} Conceptualization, Methodology, Experiment, Writing. 
{\bf Angshuman Jana:} Writing review \& editing. 
{\bf Avijit Gayen:} Supervision, Writing review \& editing.  
\end{itemize}

%\noindent\textbf{Ethical approval.} This article does not contain any studies with human
%participants or animals performed by any of the authors.
%\vspace{0.3cm}

%\noindent\textbf{Research data policy and data availability.} No exclusive data has been used for this study.                       
                             
%%===========================================================================================%%

\bibliographystyle{unsrtnat}
\bibliography{bibliography}% common bib file
%% if required, the content of .bbl file can be included here once bbl is generated
%%\input sn-article.bbl

%% Default %%
%%\input sn-sample-bib.tex%

\end{document}